\documentclass[%
 amssymb, amsmath,%
 aps,prf,superscriptaddress,longbibliography
]{revtex4-2}

\usepackage[english]{babel}

\usepackage[letterpaper,top=2.54cm,bottom=2.54cm,left=2.54cm,right=2.54cm,marginparwidth=2cm]{geometry}
\usepackage[dvipsnames,svgnames,x11names]{xcolor}
\usepackage{tikz}
\usetikzlibrary{arrows,shapes,trees,calc,positioning}
\usetikzlibrary{matrix}
\usepackage{amsmath}
\usepackage{mathrsfs}
\usepackage{graphicx}
\usepackage{subfigure}
\usepackage[colorlinks=true, allcolors=blue]{hyperref}
\usepackage{setspace}
\usepackage[rflt]{floatflt}
\usepackage{tikz}
\usetikzlibrary{decorations.pathreplacing,angles,quotes}
\usetikzlibrary{intersections}
\usepackage{amssymb}
\usepackage[titletoc]{appendix}
\usepackage{array}
\usepackage{booktabs,dcolumn,caption}
\usepackage{IEEEtrantools}
\usepackage[normalem]{ulem}

\newcommand{\We}{W\!e}
\newcommand{\mri}{\mathrm{i}}
\newcommand{\bphi}{\mbox{\boldmath$\phi$}}
\newcommand{\Bxi}{\mbox{\boldmath$\xi$}}
\newcommand{\eps}{\epsilon}
\newcommand{\bzeta}{\mbox{\boldmath$\zeta$}}

\newcommand{\bpsi}{\mbox{\boldmath$\psi$}}
\newcommand{\bxi}{\mbox{\boldmath$\xi$}}

\newcommand{\tc}{\textcolor}
\newcommand{\tcp}{\textcolor{black}}
\newcommand{\tcb}{\textcolor{black}}

\newcommand{\tcg}{\textcolor{black}}

\newcommand{\fref}[1]{Figure \ref{#1}}
\newcommand{\sref}[1]{\S~\ref{#1}}

\newcommand{\enma}[1]   {\ensuremath{#1}}

\newcommand{\non}{\nonumber}

\newcommand{\beq}{\begin{equation}}
\newcommand{\eeq}{\end{equation}}
\newcommand{\bseq}{\begin{subequations}}
\newcommand{\eseq}{\end{subequations}}
\newcommand{\beqn}{\begin{eqnarray}}
\newcommand{\eeqn}{\end{eqnarray}}
\newcommand{\ba}{\begin{array}}
\newcommand{\ea}{\end{array}}
\newcommand{\bct}{\begin{center}}
\newcommand{\ect}{\end{center}}
\newcommand{\btmz}{\begin{itemize}}
\newcommand{\etmz}{\end{itemize}}
\newcommand{\benum}{\begin{enumerate}}
\newcommand{\eenum}{\end{enumerate}}

\newcommand{\re}        {\enma{\mathrm{Re}}}
\newcommand{\im}        {\enma{\mathrm{Im}}}

\newcommand{\inner}[2]{\left\langle #1,#2 \right\rangle}

\newcommand{\be}{\begin{equation}}
\newcommand{\ee}{\end{equation}}

\newcommand{\cplxs}{ C\kern -.35em \rule{0.03 em}{.7 ex}~   }

\def\complex{\hbox{C\kern -.45em \rule{0.03 em}{1.5 ex}}~}

\newcommand{\bi}{\begin{itemize}}
\newcommand{\ei}{\end{itemize}}

\newcommand{\DefinedAs}[0]{\mathrel{\mathop:}=}
\newcommand{\AsDefined}[0]{=\mathrel{\mathop:}}

\definecolor{dkgreen}{rgb}{0,0.6,0}
\definecolor{gray}{rgb}{0.5,0.5,0.5}
\definecolor{mauve}{rgb}{0.58,0,0.82}

\newcommand{\D}[1]{\partial_{#1}}
\newcommand{\MD}{\mathrm D}
\newcommand{\DD}{\mathrm D}
\newcommand{\BB}[1]{\boldsymbol #1}

\newcommand{\R}{\mathbf R}
\newcommand{\I}{\mathbf I}
\newcommand{\0}{\mathbf 0}
\newcommand{\HH}[1]{\mathbf{#1}}
\newcommand{\MM}[1]{\mathcal{#1}}
\begin{document}

\title{Well-conditioned ultraspherical and spectral integration methods for 
	\\[0.15cm]
resolvent analysis of channel flows of Newtonian and viscoelastic fluids}

\author{Gokul Hariharan}\affiliation{Ming Hsieh Department of Electrical and Computer Engineering,\\ University of Southern California, Los Angeles, CA 90089, USA}
\author{Satish Kumar}\affiliation{Department of Chemical Engineering and Materials Science,\\ University of Minnesota, Minneapolis, MN 55455, USA}
\author{Mihailo R. Jovanovi\'c}\affiliation{Ming Hsieh Department of Electrical and Computer Engineering,\\ University of Southern California, Los Angeles, CA 90089, USA}

\newpage

\vspace*{2.cm}
\begin{abstract}
  Modal and nonmodal analyses of fluid flows provide fundamental insight into the early stages of transition to turbulence. Eigenvalues of the dynamical generator govern temporal growth or decay of individual modes, while singular values of the frequency response operator quantify the amplification of disturbances for linearly stable flows. In this paper, we develop well-conditioned ultraspherical and spectral integration methods for frequency response analysis of channel flows of Newtonian and viscoelastic fluids. Even if a discretization method is well-conditioned, we demonstrate that calculations can be erroneous if singular values are computed as the eigenvalues of a cascade connection of the frequency response operator and its adjoint. To address this issue, we utilize a feedback interconnection of the frequency response operator with its adjoint to avoid computation of inverses and facilitate robust singular value decomposition. Specifically, in contrast to conventional spectral collocation methods, the proposed method (i) produces reliable results in channel flows of viscoelastic fluids at high Weissenberg numbers ($\sim 500$); and (ii) does not require a staggered grid for the equations in primitive variables. 
  	\end{abstract}
\maketitle

	\newpage

\tableofcontents

\newpage

\section{Introduction}
	\label{sec:intro}

	\vspace*{-2ex}
Analysis of the linearized flow equations provides information about the early stages of transition to turbulence in Newtonian and viscoelastic fluids. Two broad aspects are typically considered: asymptotic stability and nonmodal amplification of disturbances in linearly stable flows~\cite{SchmidARFM2007,jovARFM20}. While eigenvalues of the linearized dynamical generator govern asymptotic growth or decay of flow fluctuations, singular values of the frequency response operator provide information about nonmodal amplification of disturbances that can trigger subcritical transition to turbulence~\cite{TrefethenS1993,butler1992three}. In particular, input-output analysis can be utilized to quantify amplification of exogenous input forcings to the linearized flow equations and provide insight into transition mechanisms~\cite{schmid2012stability,jovbamJFM05,jovARFM20}. In the presence of purely harmonic inputs, this analysis is colloquially referred to as {\em resolvent analysis\/} and the largest singular value of the frequency response operator determines the worst-case amplification of inputs with a particular temporal frequency~\cite{schmid2012stability,jovbamJFM05,jovARFM20}. This quantity determines the so-called ``resolvent norm'' and it has been used to address nonmodal amplification and robustness to modeling imperfections in channel flows of Newtonian~\cite{TrefethenS1993,mj-phd04,jovARFM20} and \mbox{viscoelastic fluids~\cite{liejovACC12,liejovkumJFM13,liejovJCP13}.} 

Resolvent analysis is typically conducted using finite-dimensional approximations of spatial differential operators in the evolution model. This model is given by a system of first-order differential equations (in time) that govern the evolution of system's state~\cite{jovARFM20}. For incompressible fluid flows, in contrast to the equations in primitive variables, the evolution model does not impose any additional constraints on the state apart from the boundary conditions. Relative to techniques based on finite-differences, pseudo-spectral collocation methods offer many advantages but may suffer from ill-conditioning of differentiation matrices~\cite{boyd1989}. Well-conditioned ultraspherical and spectral integration methods~\cite{GreSIAM91,OlvTowSIAM2013,visJCAM2015,DuSIAM2016} may alleviate these challenges while preserving convenience of traditional pseudo-spectral collocation techniques. 

In this paper, we study frequency responses of partial differential equations (PDEs) with one spatial variable and demonstrate that numerical challenges in resolvent analysis can arise even if discretization matrices of spatial differential operators are well-conditioned. We utilize the reaction-diffusion equation with homogeneous Neumann boundary conditions and strongly elastic channel flows of viscoelastic fluids to expose these challenges. For such problems, we show that reliable resolvent norm calculations can be obtained by applying well-conditioned ultraspherical and spectral integration methods to a suitable feedback interconnection of the frequency response operator with its adjoint. This formulation avoids computation of inverses and facilitates robust singular value decomposition (SVD) of the frequency response operator.

In~\cite{liejovJCP13}, the frequency response operator and its adjoint were cast as two-point boundary value problems (TPBVPs) which take the form of high-order differential equations in a spatially-independent variable. Reference~\cite{liejovJCP13} converted these TPBVPs into a system of integral equations and used Chebfun~\cite{driscoll2014chebfun} to compute the eigenvalue decomposition of a cascade connection of the frequency response operator with its adjoint. This approach avoids numerical ill-conditioning of pseudo-spectral collocation techniques in resolvent analysis and facilitates straightforward implementation of boundary conditions. 

We build on the formulation provided in~\cite{liejovJCP13} and develop a robust method for resolvent analysis of channel flows of Newtonian and viscoelastic fluids. Even if a discretization method is well-conditioned, we demonstrate that calculations can be erroneous if singular values are computed as the eigenvalues of a cascade connection of the frequency response operator and its adjoint, as done in~\cite{liejovJCP13}. To avoid computation of inverses, we utilize a feedback interconnection of the frequency response operator with its adjoint~\cite{boyd1989} and, to avoid discretization-induced ill-conditioning, we employ ultraspherical and spectral integration techniques. For incompressible fluid flows, the spectral integration method can be directly applied to the equations in primitive variables, thereby not requiring elimination of the pressure via the divergence-free condition to obtain the evolution model (which was a starting point in~\cite{liejovJCP13}). For channel flows of viscoelastic fluids, the model in primitive variables is less algebraically cumbersome than the evolution formulation~\cite{hodjovkumJFM08,harjovkumJNNFM18} and the prior work demonstrates that it yields accurate eigenvalue calculations with a smaller number of basis functions than the evolution model~\cite{khorrami1991chebyshev}. In contrast to spectral collocation techniques, our approach does not require a staggered grid and it is well-suited for resolvent analysis of strongly elastic flows of viscoelastic fluids for which the state-of-the-art approaches fail to produce reliable results. For a model in the evolution form, we also leverage Chebfun's automatic collocation technique based on ultraspherical discretization~\cite{driscoll2014chebfun}. 

Our presentation is organized as follows. In Section~\ref{sec:MotEx}, we formulate the problem and provide motivating examples that identify the need for developing well-conditioned methods for resolvent analysis. In Section~\ref{sec:numMeth}, we present a method for SVD that utilizes a feedback interconnection of the frequency response operator with its adjoint, discuss numerical methods that we employ in this work, and show how to do computations for models in primitive variables and in the evolution form. In Section~\ref{sec:apps}, we use a reaction-diffusion equation as well as channel flows of Newtonian and viscoelastic fluids to demonstrate the merits and the effectiveness of our approach. We summarize our results in Section~\ref{sec:summary} and relegate technical details to the appendix.

	\vspace*{-4ex}
\section{Problem formulation and motivating examples}
	\label{sec:MotEx}

	\vspace*{-2ex}
In this section, we formulate the problem and provide examples to motivate our developments. Our approach represents an outgrowth of the framework developed in~\cite{liejovJCP13}, where a cascade connection of the frequency response operator and its adjoint was utilized in nonmodal analysis of stable linear dynamical systems in which the spatial variable belongs to a finite interval.

	\vspace*{-8ex}
\subsection{Problem formulation}\label{sec:probForm}
  
  \vspace*{-2ex}
We consider linear dynamical systems whose spatio-temporal frequency response $\MM T(\omega)$ can be cast as,  	\begin{subequations}\label{eq:mot1}
\begin{align}
    \left[\MM A(\omega)\,\bphi(\cdot)\right] (y) \;&=\; \left[\MM B(\omega)\,\BB d(\cdot)\right](y),\\
    \bxi (y)\;&=\; \left[\MM C(\omega)\,\bphi(\cdot)\right](y),\\
    [\MM{L}_a \, \bphi(\cdot)](a)  \;&=\; [\MM{L}_b \, \bphi(\cdot)](b) \;=\;  0,\label{eq:mot1c}
  \end{align}
  \end{subequations}
 where $\omega \in \mathbb{R}$ is the temporal frequency and $y \in [a, b]$ is the spatial variable. The state, input, and output fields are respectively denoted by $\bphi$, $\BB{d}$, and $\Bxi$; $\MM A$, $\MM{B}$, and $\MM{C}$ are linear differential block matrix operators of appropriate dimensions with potentially non-constant coefficients in $y$; and $\MM{L}_a$ and $\MM{L}_b$ are linear operators that specify the boundary conditions on $\bphi$. At any temporal frequency, we assume that the operator $\MM A(\omega)$ in~\eqref{eq:mot1} is invertible, thereby leading to,
	\be
	\MM T(\omega)
	\; = \;
	\MM{C}(\omega) \MM A^{-1}(\omega) \MM{B}(\omega).
	\non
	\ee
While we allow a nonlinear dependence of the operators $\MM A$, $\MM B$, and $\MM C$ on $\omega$, for systems that can be cast as,
    \begin{subequations}\label{eq:mot1E}
  \begin{alignat}{3}
  \D{t} [\MM{E} \, \bphi( \cdot , t)](y) \;&=\; [\MM{F}\, \bphi(\cdot , t)](y) \,+\, [\MM{B}\,\BB{d}(\cdot,t)](y),&&\label{eq:mot1Ea}\\
    \Bxi(y,t) \;&=\; [\MM{C}\,\bphi(\cdot,t)](y),&\label{eq:mot1Eb}\\
   [\MM{L}_a \, \bphi(\cdot,t)](a)  \;&=\; [\MM{L}_b \, \bphi(\cdot,t)](b) \;=\;  0,&\label{eq:mot1Ec}
\end{alignat}
\end{subequations}
the operator $\MM A (\omega)$ in~\eqref{eq:mot1} depends linearly on $\omega$, where $t\in [0, \infty)$ is time. In this case, the application of the temporal Fourier transform yields the resolvent operator, $\MM A^{-1}(\omega) = (\mathrm{i} \omega \MM{E} - \MM{F})^{-1}$, where $\mri$ is the imaginary unit, and the operators $\MM B$ and $\MM C$ in~\eqref{eq:mot1} do not depend on $\omega$.

The frequency response operator ${\MM T}(\omega)$ determines the steady-state response of a stable linear dynamical system to  purely harmonic inputs. Namely, for $\BB{d}(y,t) = \hat{\BB{d}} (y,\omega) \mathrm{e}^{\mri \omega t}$, the steady-state response is given by $\Bxi(y,t) = \hat{\Bxi} (y,\omega) \mathrm{e}^{\mri \omega t}$ and ${\MM T}(\omega)$ maps a spatial input profile $\hat{\BB{d}} (y,\omega)$ into the corresponding output $\hat{\Bxi} (y,\omega)$,
	\be
	\hat{\Bxi} (y,\omega)
	\; = \;
	\left[
	\MM T(\omega) \, \hat{\BB{d}} (\cdot,\omega)
	\right]
	(y).
	\non
  \ee
The singular value decomposition (SVD) of $\MM T(\omega)$ can be used to determine the input shapes (i.e., the left singular functions $\hat{\BB{v}}_i (y,\omega)$), the resulting responses (i.e., the right singular functions $\hat{\BB{u}}_i (y,\omega)$), and the corresponding gains (i.e., the singular values $\sigma_i (\omega) $),
	\be
	\hat{\Bxi} (y,\omega)
	\; = \;
	\left[
	\MM T(\omega) \, \hat{\BB{d}} (\cdot,\omega)
	\right]
	(y)
	\; = \;
	\sum_{i \, = \, 0}^{\infty}
	\sigma_i (\omega)
	\hat{\BB{u}}_i (y,\omega)
	\langle {\hat{\BB{v}}_i (\cdot,\omega)}, {\hat{\BB{d}} (\cdot,\omega)} \rangle,
	\non
	\ee
where $\inner{\cdot}{\cdot}$ is the standard $L_2 [a, b]$ inner product. SVD requires computation of the adjoint $\MM T^{\dagger}(\omega)$ of $\MM T(\omega)$,
	\be
	\langle \MM T^{\dagger} (\omega) \hat{\Bxi}, \hat{\BB{d}} \rangle
	\; = \;
	\langle \hat{\Bxi}, \MM T (\omega)  \hat{\BB{d}} \rangle,
	\non
	\ee
and the eigenvalue decomposition of the composite operators $\MM T(\omega)\MM T^{\dagger}(\omega)$ and $\MM T^{\dagger}(\omega) \MM T(\omega)$~\cite{liejovJCP13,schmid2012stability},
	\be
	\ba{rclrcl}
	\left[
	\MM T(\omega) \MM T^{\dagger}(\omega)
	\hat{\BB{u}}_i (\cdot,\omega)
	\right] (y)
	& \! = \! &
	\sigma_i^2 (\omega)
	\hat{\BB{u}}_i (y,\omega),
	~~
	\left[
	\MM T^{\dagger} (\omega) \MM T (\omega)
	\hat{\BB{v}}_i (\cdot,\omega)
	\right] (y)
	& \! = \! &
	\sigma_i^2 (\omega)
	\hat{\BB{v}}_i (y,\omega).
	\ea
	\non
	\ee
\tcb{Here, $\MM T^{\dagger}(\omega)$ is the continuous adjoint which} is not determined by the complex conjugate transpose of an operator-valued matrix $\MM T (\omega)$; its computation typically involves integration by parts \tcb{and discretization of the resulting equation.}

	\vspace*{-4ex}
\subsection{Reaction-diffusion equation with Neumann boundary conditions}\label{sec:MotEx1}

	\vspace*{-2ex}
For the reaction-diffusion equation with $y\in [-1, 1]$ and homogeneous Neumann boundary conditions,
	\begin{align}\label{eq:rnd}
\begin{split}
\partial_t \phi(y,t) \;&=\; \partial_{yy} \phi (y,t)  \,-\, \epsilon^2 \phi (y,t)  \,+\,  d (y,t),
	\\[0.1cm]
\partial_y \phi (\pm 1,t)   \;&=\;0,
\end{split}
\end{align}
where $\epsilon$ is a real parameter, in representation~\eqref{eq:mot1} we have
	$$
	\MM A (\omega)\;=\; \mri \omega I \,-\, \MD^2 \, + \, \epsilon^2 I,
	~~
	\MM B \;=\; \MM C \;=\; I.
	$$
Here, $I$ is the identity operator, $\MD = \mathrm d / \mathrm dy$, and the frequency response operator is determined by
\begin{align}\label{eq:rnd_frop}
  \MM{T}(\omega)  \;=\; \left( \mri \omega I \,-\, \MD^2 \, + \, \epsilon^2 I \right)^{\,-1}.
\end{align}
 The dynamical generator $\MD^2 - \epsilon^2 I$ with homogeneous Neumann boundary conditions in~\eqref{eq:rnd} is self-adjoint and its eigen-pairs are given by~\cite[Example 5.4-1]{deen1998analysis}
	\begin{align}\label{eq:rnd_analLam}
	\begin{array}{rclccl}
	\lambda_n &=& - ( \eps^2 + n^2\pi^2 ),& \phi_n(y) &=& \cos \left( n\pi y \right),
	\\[0.15cm]
	\lambda_n &=& -( \eps^2 + ( n+ \tfrac{1}{2} )^2 \pi^2 ),& \phi_n(y) &=& \sin \left(  ( n+\tfrac{1}{2} ) \pi y \right),
	\end{array}
	\end{align}
where $n\in \mathbb Z$. Furthermore, the singular values of the frequency response operator are determined by
	\begin{align}\label{eq:rnd_anal}
\sigma_n^2 (\omega) \;&=\;
\begin{cases}
  \; 1/{(\omega^2 + (\epsilon^2 + n^2\,\pi^2)^2)},
  \\[0.15cm]
  \; 1/{(\omega^2 + (\epsilon^2 +  ( n+\tfrac{1}{2} )^2 \pi^2)^2)},
\end{cases}
\end{align}
and the largest value of $\sigma_n (\omega)$ occurs for $n = 0$ and $\omega = 0$, i.e., $\max_{n, \, \omega} \sigma_n (\omega) = \sigma_0 (0) = 1/\epsilon^2$.

\begin{figure}
\centering
\subfigure[][$\epsilon = 1$, $\omega = 0$]{\includegraphics[scale = 0.21]{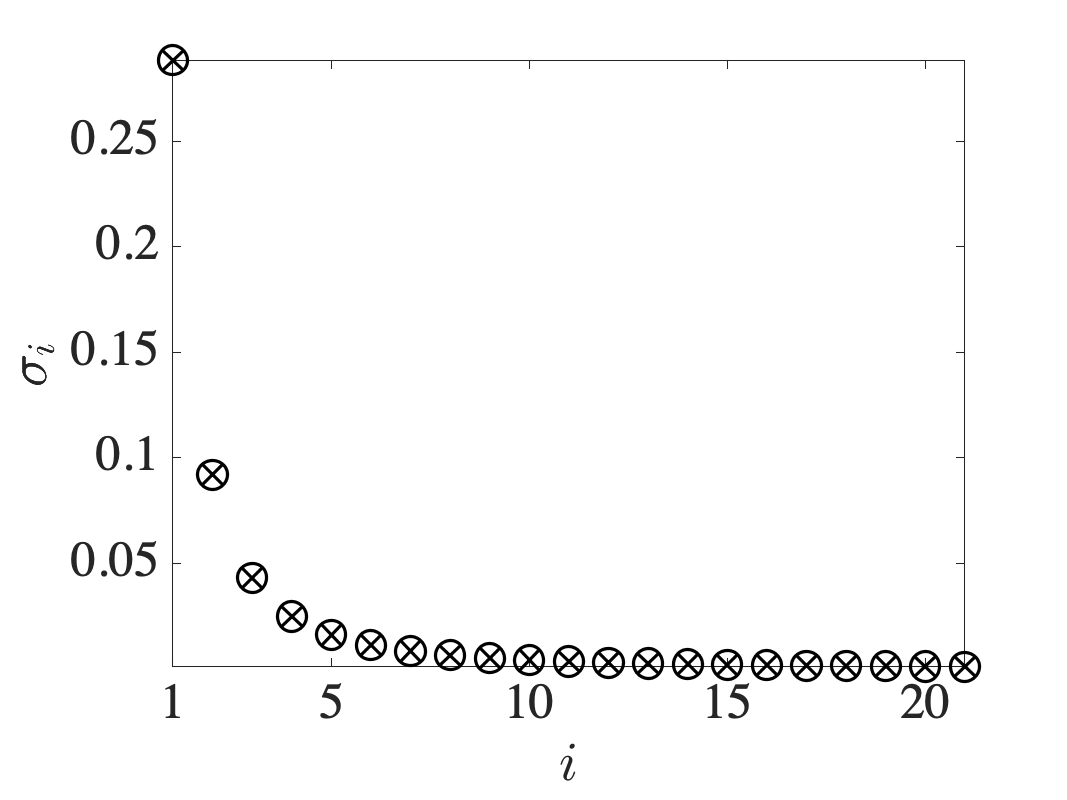}\label{fig:1a}}
\subfigure[][$\epsilon = 10^{-4}$, $\omega = 0$]{\includegraphics[scale = 0.21]{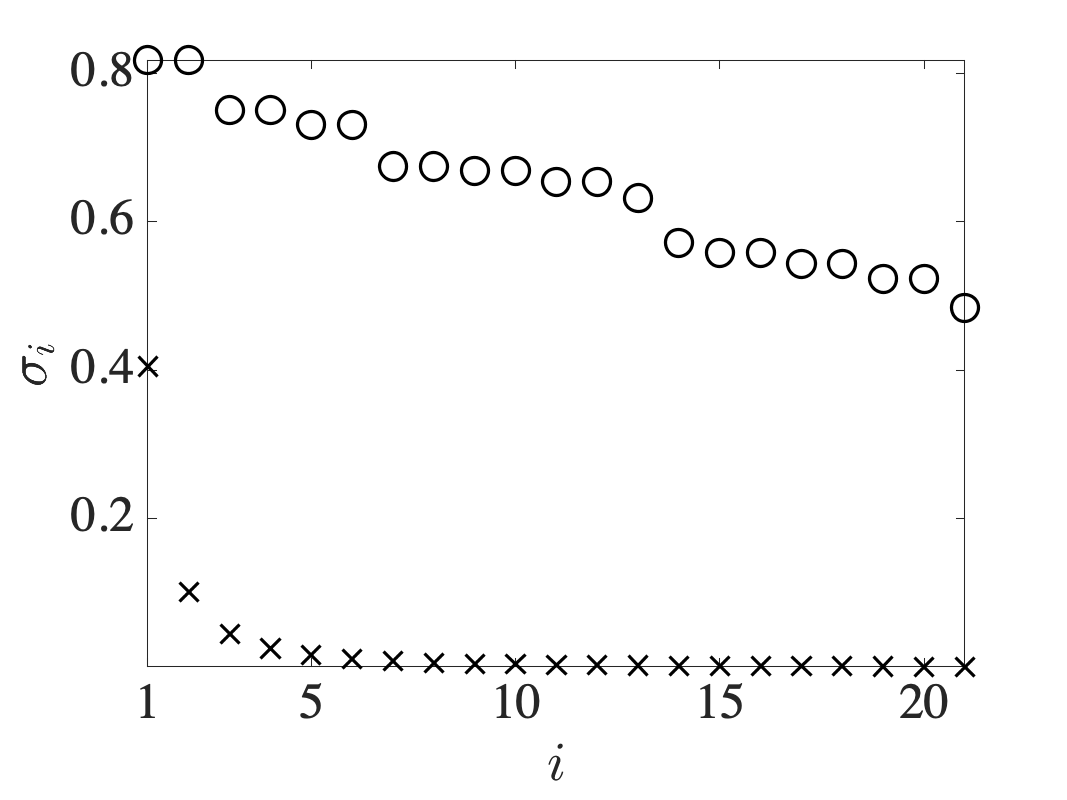}\label{fig:1b}}
  \caption{\label{fig:1} Singular values of the frequency response operator of the reaction-diffusion equation \eqref{eq:rnd} obtained using Chebfun's spectral scheme with $N = 64$ collocation points. Symbols represent exact values ($\times$) and the numerical solution resulting from the composite operator $\MM T(\omega)\MM T^{\dagger}(\omega)$ ($\circ$). The principal singular value is not shown as its value is very large compared to the remaining singular values.}
  	\vspace*{-0.25cm}
\end{figure}

The separation between $\sigma_0 (0)$ and $\sigma_1 (0)$ increases with decrease in $\epsilon$ and this ill-conditioning negatively impacts performance of standard numerical schemes. Singular value decomposition typically involves the resolvent operator and its numerical evaluation requires computation of the inverse of the discretized version of an operator-valued matrix. \fref{fig:1} illustrates that computations based on the composite operator $\MM T(\omega)\MM T^{\dagger}(\omega)$ yield erroneous results for reaction-diffusion equation~\eqref{eq:rnd} with small values of $\epsilon$. The collocation method with $64$ Chebyshev basis functions is used and similar results are obtained even with a well-conditioned spectral integration scheme. For $\epsilon = 10^{-4}$, $\sigma_0 (0) = 10^8$ is significantly larger than the other singular values and it is not shown in Figure~\ref{fig:1}. \tcg{The largest singular value is computed correctly to a relative accuracy of $\mathcal O (10^{-5})$ using both approaches.} Even though the collocation method is well-conditioned for $64$ basis functions~\cite{boy2001}, singular values resulting from spatial discretization of the composite operator $\MM T(\omega)\MM T^{\dagger}(\omega)$ have non-zero imaginary parts and their real parts significantly deviate from the true values; see Figure~\ref{fig:1b}. When the composite operator is used, increasing the number of basis functions does not fix this problem. In contrast, for $\epsilon = 1$ (\fref{fig:1a}), we observe a good match between analytical solutions (marked by crosses) and singular values calculated using the composite operator $\MM T(\omega)\MM T^{\dagger}(\omega)$ (marked by circles). 
 
In this example, since $\MM B = \MM C = I$ and $\MM T^\dagger (\omega) \MM T(\omega) = \MM A^{-\dagger}(\omega)\MM A^{-1}(\omega) = ( \MM A(\omega)\MM A^{\dagger}(\omega) )^{-1}$, ill-conditioning can be circumvented by computing the eigenvalues of the operator $\MM A(\omega)\MM A^{\dagger}(\omega)$. However, in general, $\MM B$ and $\MM C$ are nonsquare block-matrix operators and the computation of $\MM A^{-1} (\omega)$ and ${\MM A}^{-\dagger}(\omega)$ cannot be avoided when a cascade connection of $\MM T(\omega)$ and $\MM T^\dagger (\omega)$, shown in~\fref{fig:cascade}, is used in the frequency response analysis. As described in \S~\ref{sec:apps2}, similar operator-induced ill-conditioning arises in strongly elastic flows of viscoelastic fluids. \tcg{Furthermore, reaction-diffusion system~\eqref{eq:rnd} is normal and the largest singular value of the frequency response operator is reliably captured by both cascade and feedback interconnections with similar level of accuracy. On the other hand, for a cascade interconnection shown in~\fref{fig:1b} all other singular values are erroneous and they contain large imaginary parts. For non-normal systems, e.g., viscoelastic channel flows, the largest singular value can also have significant imaginary part. A framework that avoids matrix inversions can potentially deal with challenging cases in which a system is close to being singular.} 

In \S~\ref{sec:feedback} and \S~\ref{sec:apps1}, we revisit the reaction-diffusion problem and show that the use of a feedback interconnection, shown in \fref{fig:cas2feedback}, leads to a computational framework that is insensitive to ill-conditioning of the underlying operator.  \tcb{While using a higher precision arithmetic (e.g., quadruple) may mitigate errors associated with inversion of matrices~\cite{golub2012matrix}, required level of precision would depend on the value of the reaction rate. In contrast, the feedback interconnection shown in~\fref{fig:cas2feedback} is insensitive to such ill-conditioning.} 

	\vspace*{-4ex}
\subsection{Channel flow of viscoelastic fluids}
	\label{subsec:illcondDis}

	\vspace*{-2ex}
We now examine the model that governs the dynamics of infinitesimal fluctuations around the laminar flow of a dilute polymer solution in a channel. This problem was used in~\cite{liejovJCP13} to demonstrate that spectral collocation and an integral reformulation of spectral collocation can produce significantly different results with accurate and grid-independent results only feasible with the latter. In \S~\ref{sec:apps2}, we show that ultraspherical discretization offers a similar level of accuracy as spectral integration and that under similar conditions, spectral collocation performs poorly, which is in concert with the observations made in~\cite{liejovJCP13}.

The linearized momentum, mass conservation, and constitutive equations for an incompressible flow of the Oldroyd-B fluid are given by~\cite{harjovkumJNNFM18,jovkumPOF10,jovkumJNNFM11,hodjovkumJFM08,hodjovkumJFM09},
	\begin{subequations}\label{eq:2.1}
\begin{align}
  Re \;\! ( \partial_t \BB{v}  \, + \,  \BB{V} \cdot \BB{\nabla}\BB{v}  \, + \,  \BB v \cdot \BB{\nabla}\BB{V} )     
  \;=&~\, 
  - \BB{\nabla}p  \, + \,   \beta\, \nabla^2 \BB{v}  \, + \,  (1-\beta)\, \BB{\nabla}  \cdot \BB{\tau} \, + \,  \BB{d}, 
  \label{eq:2.1a}
  	\\[0.1cm]
    \BB{\nabla} \cdot \BB{v}  \;=&~\, 0,
    \label{eq:2.1b}
	\\[0.1cm]  
	\begin{split}  
	\partial_t\BB{\tau} \, + \, \BB V \cdot \BB{\nabla}\BB{\tau}  \, + \,  \BB{v} \cdot \BB{\nabla} \BB{\tau} \;=&~\,    \BB{\tau} \cdot \BB{\nabla}\BB{V} \, + \,    \left( \BB{\tau} \cdot \BB{\nabla}\BB{V} \right) ^T  \, + \,  \BB{T} \cdot \BB{\nabla}\BB{v} \, + \,   \left( \BB{T} \cdot \BB{\nabla}\BB{v} \right) ^T 
	~ +
	\\ \;&~\,
         \frac{1}{\We} \left(\BB{\nabla}\BB{v}  \, + \,  \BB{\nabla}\BB{v}^T  \, - \, \BB{\tau} \right).
    \end{split} \label{eq:2.1c}
\end{align}
\end{subequations}
Here, $\BB v$, $\BB \tau$, and $p$ are velocity, stress, and pressure fluctuations around the corresponding base-flow quantities $\BB V$, $\BB T$, and $P$, respectively. The length is normalized with the half-channel height $h$ (see \fref{fig:flow_geo} for geometry), velocity with the largest value of the steady-state velocity $U_0$, time with $h/U_0$, pressure with $\mu_TU_0/h$ where $\mu_T$ is the effective shear viscosity of the dilute viscoelastic solution, and the polymer stress $\BB T$ with $\mu_p U_0 /h$, where $\mu_p = \mu_T - \mu_s$ and $\mu_s$ is the pure-solvent viscosity. The Reynolds number, $Re = h U_0 \rho / \mu_{T}$, quantifies the ratio between the inertial and viscous forces, where $\rho$ is the fluid density; the Weissenberg number, $\We = \lambda_p U_0 /h$, provides a measure of the degree of elasticity in the fluid, where $\lambda_p$ is the polymer relaxation time; and the viscosity ratio, $\beta = \mu_{s}/\mu_T$, determines the polymer concentration in the fluid. Setting $\beta = 0$ in~\eqref{eq:2.1} yields an upper convected Maxwell (UCM) model and for $\beta = 1$ a flow of Newtonian fluid is recovered.

\setlength{\unitlength}{0.6cm}
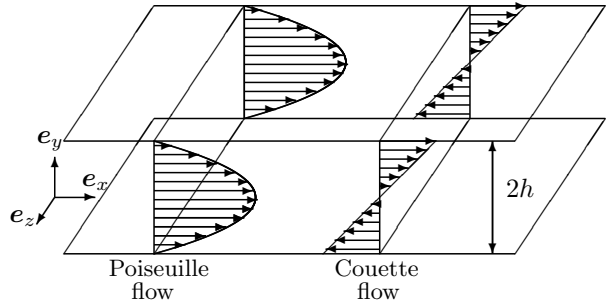
\begin{figure}
\begin{picture}(21,7)(-4,-1)
\linethickness{0.2mm}
\multiput(1,0)(4,0){1}%
{\line(1,0){10}}
\multiput(3,0)(4,0){1}%
{\line(0,1){2.5}}
\multiput(1,2.5)(4,0){1}%
{\line(1,0){10}}
\multiput(0.8,1.25)(4,0){1}%
{\vector(0,1){0.9}}
\multiput(0.8,1.25)(4,0){1}%
{\vector(1,0){0.9}}
\multiput(0.8,1.25)(4,0){1}%
{\vector(-2,-3){0.4}}
\put(0.4,2.45){$\BB e_y$}
\put(1.4,1.5){$\BB e_x$}
\put(-0.2,0.7){$\BB e_z$}
\put(2.0,-0.5){\small Poiseuille}
\put(2.5,-1){\small flow}
\put(7.0,-0.5){\small Couette}
\put(7.5,-1){\small flow}
\qbezier(3,0)(7.5,1.25)(3,2.5)
 \put(3,0.2){\vector(1,0){0.7}}
 \put(3,0.4){\vector(1,0){1.15}}
 \put(3,0.6){\vector(1,0){1.6}}
 \put(3,0.8){\vector(1,0){2}}
 \put(3,1){\vector(1,0){2.2}}
 \put(3,1.2){\vector(1,0){2.3}}
 \put(3,1.4){\vector(1,0){2.2}}
 \put(3,1.6){\vector(1,0){2.1}}
 \put(3,1.8){\vector(1,0){1.8}}
 \put(3,2){\vector(1,0){1.4}}
 \put(3,2.2){\vector(1,0){1}}
 \put(3,2.4){\vector(1,0){0.4}}
 \qbezier(5,3)(9.5,4.25)(5,5.5)
 \put(5,3.2){\vector(1,0){0.7}}
 \put(5,3.4){\vector(1,0){1.15}}
 \put(5,3.6){\vector(1,0){1.6}}
 \put(5,3.8){\vector(1,0){2}}
 \put(5,4){\vector(1,0){2.2}}
 \put(5,4.2){\vector(1,0){2.3}}
 \put(5,4.4){\vector(1,0){2.2}}
 \put(5,4.6){\vector(1,0){2.1}}
 \put(5,4.8){\vector(1,0){1.8}}
 \put(5,5){\vector(1,0){1.4}}
 \put(5,5.2){\vector(1,0){1}}
 \put(5,5.4){\vector(1,0){0.4}}
\put(10.5,0){\vector(0,1){2.5}}
\put(10.5,2.5){\vector(0,-1){2.5}}
\put(10.8,1.25){$2h$}
\put(3,2.5){\line(2,3){2}}
\put(3,0){\line(2,3){2}}
\put(5,3){\line(0,1){2.5}}
\put(5,3){\line(0,1){2.5}}
\put(1,2.5){\line(2,3){2}}
\put(11,2.5){\line(2,3){2}}
\put(3,5.5){\line(1,0){10}}
\put(11,0){\line(2,3){2}}
\put(1,0){\line(2,3){2}}
\put(3,3){\line(1,0){10}}
\put(8,2.5){\line(2,3){2}}
\put(8,0){\line(2,3){2}}
\put(10,3){\line(0,1){2.5}}
\put(8,0){\line(0,1){2.5}}
\put(6.75,0){\line(1,1){2.5}}
\put(8.75,3){\line(1,1){2.5}}
\put(8,0.1){\vector(-1,0){1.2}}
\put(8,0.3){\vector(-1,0){1.0}}
\put(8,0.5){\vector(-1,0){0.8}}
\put(8,0.7){\vector(-1,0){0.6}}
\put(8,0.9){\vector(-1,0){0.4}}
\put(8,1.1){\vector(-1,0){0.2}}
\put(8,1.5){\vector(1,0){0.25}}
\put(8,1.7){\vector(1,0){0.45}}
\put(8,1.9){\vector(1,0){0.65}}
\put(8,2.1){\vector(1,0){0.85}}
\put(8,2.3){\vector(1,0){1.05}}
\put(8,2.5){\vector(1,0){1.25}}

\put(10,3.1){\vector(-1,0){1.2}}
\put(10,3.3){\vector(-1,0){1.0}}
\put(10,3.5){\vector(-1,0){0.8}}
\put(10,3.7){\vector(-1,0){0.6}}
\put(10,3.9){\vector(-1,0){0.4}}
\put(10,4.1){\vector(-1,0){0.2}}
\put(10,4.5){\vector(1,0){0.25}}
\put(10,4.7){\vector(1,0){0.45}}
\put(10,4.9){\vector(1,0){0.65}}
\put(10,5.1){\vector(1,0){0.85}}
\put(10,5.3){\vector(1,0){1.05}}
\put(10,5.5){\vector(1,0){1.25}}
 \put(3,0.4){\vector(1,0){1.15}}
 \put(3,0.6){\vector(1,0){1.6}}
 \put(3,0.8){\vector(1,0){2}}
 \put(3,1){\vector(1,0){2.2}}
\end{picture}
\caption{\label{fig:flow_geo} Geometry and steady-state velocity profiles in Poiseuille and Couette flows.}
	\vspace*{-0.5cm}
\end{figure}

In channel flow, the steady-state velocity profile only contains the streamwise component, i.e., $\BB V = (U(y),0,0)$, where $U(y) = 1-y^2$ for pressure-driven Poiseuille flow and $U(y) = y$ for shear-driven Couette flow. The non-zero components of the base stress tensor are given by $T_{xx} = 2 \We ( U'(y) )^2$ and $T_{xy} = T_{yx} = U'(y)$, where the prime denotes a derivative with respect to $y$. For this base flow, the streamwise and spanwise directions are translationally invariant and the spatio-temporal Fourier transform brings~\eqref{eq:2.1} to a two-point boundary value problem in the wall-normal coordinate $y$.

In the absence of inertia, we can set $Re = 0$ in~\eqref{eq:2.1}, rescale time with $\We$, and examine the dynamics of 2D velocity fluctuations $\BB v = (u, v)$ in the streamwise/wall-normal plane $(x,y)$. Introducing the streamfunction $\phi$ so that the streamwise and wall-normal velocity components are given by $u = \partial_y \phi$ and $v = -i k_x \phi$ and eliminating pressure and stress fluctuations from~\eqref{eq:2.1} brings the frequency response operator $\MM T(\omega)$ into the following form with $\MD = \mathrm d / \mathrm dy$, 
	\begin{subequations}
\begin{equation}\label{eq:2Dvisco}
  \begin{aligned}
   \left(   \sum_{n \, = \, 0}^4 a_n(y,\omega)\MD^n   \right)     \phi (y,\omega)  \;&=\;
 \left[ \begin{array}{cc}
   \MD & -\mri k_x
 \end{array} \right]
 \left[ \begin{array}{c}
  d_x (y,\omega) \\
  d_y (y,\omega)
 \end{array} \right],\\
 \left[ \begin{array}{c}
  u (y,\omega) \\
  v (y,\omega)
 \end{array} \right]
 \;&=\;
 \left[ \begin{array}{c}
  \MD
  \\
  -\mri k_x
 \end{array} \right] \phi (y,\omega),
 \\[0.1cm]
 \phi (\pm 1,\omega) \;&=\; [ \MD  \phi (\cdot,\omega) ] (\pm 1) \;=\; 0,
\end{aligned}
\end{equation}
thereby implying that, in representation~\eqref{eq:mot1}, we have,
\begin{align*}
  \MM A(\omega) \;=\; \sum_{n \, = \, 0}^4 a_n(y,\omega)\MD^n,  
  ~~
  \MM B \;=\; \left[ \begin{array}{cc}
    \MD & -\mri k_x
  \end{array} \right],
  	~~
  \MM C \;&=\;\left[ \begin{array}{c}
    \MD
    \\
    -\mri k_x
   \end{array} \right].
\end{align*}
Alternatively, the components of the fluctuation stress tensor, which can play an active role in triggering instabilities in viscoelastic fluids~\cite{larson1999structure}, can be selected as the output in~\eqref{eq:2Dvisco},
\begin{align}\label{eq:stressout}
   \left[ \begin{array}{c}
     \tau_{xx} (y,\omega) \\
     \tau_{xy} (y,\omega) \\
     \tau_{yy} (y,\omega)
   \end{array} \right] 
   \; =  \; 
   \left[ \begin{array}{c}
     c_{11}(y,\omega)\MD^2 \, + \, c_{12}(y,\omega)\MD \, + \, c_{13}(y,\omega) \\
     c_{21}(y,\omega)\MD^2 \, + \, c_{22}(y,\omega)\MD \, + \, c_{23}(y,\omega) \\
     c_{31}(y,\omega)\MD \, + \, c_{32}(y,\omega)\!\\
   \end{array} \right] \phi (y,\omega).
\end{align}
\end{subequations}
The expressions for functions $a_n(y,\omega)$ and $c_{ij}(y,\omega)$ are provided in Appendix~\ref{appA}.

	\vspace*{-4ex}
\subsection{The linearized Navier-Stokes equations: A model in the descriptor form} 
	\label{sec:MotEx3}

	\vspace*{-2ex}
  Setting $\beta = 1$ and rescaling pressure with $Re$ in~\eqref{eq:2.1} yields the linearized Navier-Stokes (NS) equations,
\begin{subequations}\label{eq:lns}
\begin{align}
   \,\partial_t \BB{v}  \,+\, \BB{V} \cdot \BB{\nabla}\BB{v}  \,+\, \BB v \cdot \BB{\nabla}\BB{V} \;=&~\, - \BB{\nabla}p  \,+\,  \frac{1}{Re}\,\nabla^2 \BB{v} \,+\, \BB{d}, \label{eq:lnsa}
   	\\
   \BB{\nabla} \cdot \BB{v}  \;= &~\, 0. \label{eq:lnsb}
\end{align}
\end{subequations}
At any time $t$, the velocity fluctuations in~\eqref{eq:lns} have to satisfy the algebraic constraint given by the continuity equation~\eqref{eq:lnsb}. In channel flow, the application of the Fourier transform in $x$, $z$, and $t$ allows us to cast~\eqref{eq:lns} in the form given by~\eqref{eq:mot1} which is parameterized by the wall-parallel wavenumbers ($k_x,k_z$) and the temporal frequency $\omega$. Using a standard procedure~\cite[Chapter 3]{schmid2012stability}, pressure can be eliminated from~\eqref{eq:lns} to obtain a model in the evolution form in which the state is captured by the wall-normal velocity and vorticity fluctuations, ($v,\eta$). When the pressure is kept in the governing equations, we deal with a model in the {\em descriptor form\/} in which the state is captured by the primitive variables ($u,v,w,p$). 

Bringing~\eqref{eq:lns} to the evolution form has advantages and disadvantages. This transformation eliminates the need to deal with pressure boundary conditions, which are unknown, and it yields a smaller number of state variables. However, there are considerable disadvantages both in Newtonian and viscoelastic fluids. As shown in~\cite{khorrami1991chebyshev,KHORRAMI1989206}, for the same level of accuracy, the descriptor form in channel flows of Newtonian fluids requires a smaller number of basis functions compared to the evolution form. Furthermore, in flows of viscoelastic fluids, the transformation to the evolution form can result in a system that is algebraically cumbersome (e.g., see Appendices in~\cite{harjovkumJNNFM18,zakiJFM2013}) and eliminating pressure from~\eqref{eq:2.1} requires taking higher derivatives of the stress variables and necessitates specification of additional boundary conditions on stress fluctuations. Certain boundary conditions on stress fluctuations have been identified to produce reliable results~\cite{GraJFM1998}, but the physical basis of these remains unclear.

Since the boundary conditions on pressure are not known, working with the model in the descriptor form requires use of a staggered grid for the velocity and pressure fields in the spectral collocation method. If velocity is evaluated at Chebyshev collocation points,
	\begin{subequations}
\begin{align}
  y_j \, = \, \cos  \left( {\pi j}/{N} \right),
  \quad j \, = \, 0,1,\ldots, N,
\end{align}
then the pressure is evaluated at the points
\begin{align}
	\label{eq:presNodes}
  y_j \, = \, \cos \left( {\pi  ( j \, + \, \tfrac{1}{2} ) }/{N} \right),
  \quad j \, = \, 0,1,\ldots, N-1;
\end{align}
	\end{subequations}
when using a staggered grid. A similar procedure for the Chebyshev-Tau method is described in \cite{CanHusQuaTho2012}. By setting $j = 0$ and $j = N-1$ in~\eqref{eq:presNodes}, pressure is not evaluated at the boundaries, i.e., at $ y = \pm 1$, and thus the need for specifying pressure boundary conditions is avoided. We refer the reader to~\cite{gotorsSIAM1997} for implementation details of the staggered-grid formulation.

However, implementing a staggered grid can be challenging and there are well-developed open-source codes to solve two-point boundary value problems using spectral methods, e.g., A Matlab Differentiation Matrix Suite~\cite{weideman2000matlab} and Chebfun~\cite{driscoll2014chebfun}. Implementing staggered grids in such solvers requires special treatment and the standard solvers currently available in Chebfun do not cater to unconventional discretizations. In \S~\ref{sec:apps3}, we demonstrate that the Chebyshev spectral integration method does not need a staggered grid when retaining the problem in the descriptor form and reinforcing algebraic constraint~\eqref{eq:lnsb} at the boundaries, $y = \pm 1$.

In channel flow of a viscoelastic fluid, the momentum equation in~\eqref{eq:2.1} contains the divergence of stress fluctuations and the presence of the  $y$-derivative of $\BB{\tau}$ complicates determination of boundary conditions for the adjoint system. In~\S~\ref{sec:desStressElim}, we develop a method for resolvent analysis that retains the accuracy of the descriptor formulation and circumvents the challenge of dealing with stress boundary conditions. In our approach, we eliminate the stress fluctuations from~\eqref{eq:2.1}, while retaining the pressure, and exploit the fact that the spectral integration method does not require a staggered grid when pressure is kept in the governing equations. In~\sref{sec:app3b}, we demonstrate that our spectral integration implementation of the descriptor formulation provides a reliable tool for conducting the frequency response analysis in 3D channel flow of a viscoelastic fluid even in strongly elastic regimes. 

\vspace*{-4ex}
\section{Singular value decomposition via feedback interconnection}
	\label{sec:numMeth}

\begin{figure}
\centering
\begin{tabular}{c}
%
%
%
%
\input{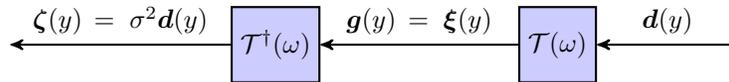}
%
%
\noindent
\begin{tikzpicture}[scale=1, auto, >=stealth']

	\node[] (zetaend) at (0,0) {};
		 
    \node[block, minimum height = 1.cm, top color=blue!20, bottom color=blue!20] (Tstar) at ($(zetaend) + (3.65cm,0)$) 
    {$\MM{T}^{\dagger}(\omega)$};
    
    \node[block, minimum height = 1.cm, top color=blue!20, bottom color=blue!20] (T) at ($(Tstar) + (3.75cm,0cm)$) 
    {$\MM{T} (\omega)$};

     	\node[] (dbegin) at ($(T.east) + (2.cm,0)$) {};
	
    	\draw [connector] (dbegin.west) -- node [midway, above] {$\BB d(y)$} (T.east);
    
    	\draw [connector] (T.west) -- node [midway, above] {$\BB g (y) \, = \; \Bxi (y)$} (Tstar.east);
	
    	\draw [connector] (Tstar.west) -- node [midway, above] {$\bzeta(y) \, = \; \sigma^2 \BB d (y)$} (zetaend);
	
\end{tikzpicture}
\end{tabular}
  \caption{\label{fig:cascade} Block diagram of a cascade connection of the operators $\MM{T}^{\dagger}(\omega)$ and $\MM T(\omega)$. The composite operator, $\MM{T}^{\dagger}(\omega) \MM T(\omega)$, can be used to compute the singular values of the frequency response operator $\MM T(\omega)$.}
\end{figure}

\begin{figure}
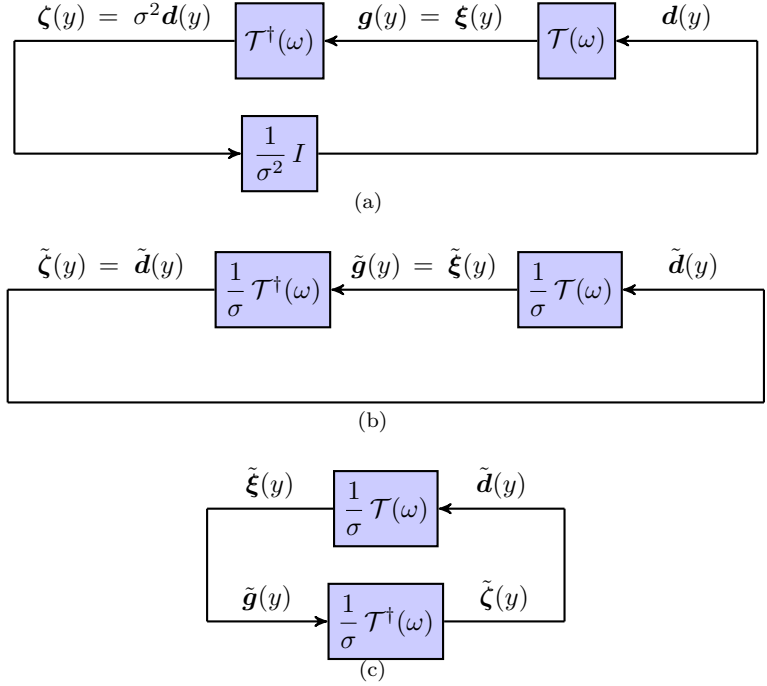

\centering
\begin{tabular}{c}
\begin{tabular}{c}
    \subfigure[]{
%
%
%
%
\input{tikz_common_styles}
%
%
\noindent
\begin{tikzpicture}[scale=1, auto, >=stealth']

	\node[] (zetaend) at (0,0) {};
	
	 \node[] (fend) at ($(zetaend.center) - (0.cm,1.5cm)$) {};
		 
    \node[block, minimum height = 1.cm, top color=blue!20, bottom color=blue!20] (Tstar) at ($(zetaend) + (3.65cm,0)$) 
    {$\MM{T}^{\dagger}(\omega)$};
    
    \node[block, minimum height = 1.cm, top color=blue!20, bottom color=blue!20] (T) at ($(Tstar) + (3.95cm,0cm)$) 
    {$\MM{T} (\omega)$};
    
    \node[block, minimum height = 1.cm, top color=blue!20, bottom color=blue!20] (fbk) at ($(fend) + (3.65cm,0)$) 
    {$\dfrac{1}{\sigma^2} \, I$};

     	\node[] (dbegin) at ($(T.east) + (2.cm,0)$) {};
	
     	\node[] (fbegin) at ($(dbegin.west) - (0.cm,1.5cm)$) {};
	
    	\draw [connector] (dbegin.west) -- node [midway, above] {$\BB d(y)$} (T.east);
    
    	\draw [connector] (T.west) -- node [midway, above] {$\BB g (y) \, = \; \Bxi (y)$} (Tstar.east);
	
    	\draw [line] (Tstar.west) -- node [midway, above] {$\bzeta(y) \, = \; \sigma^2 \BB d (y)$} (zetaend);
	
    	\draw [line] (zetaend.east) -- node [midway, above] {$$} (fend.east);
	
    	\draw [connector] (fend.east) -- node [midway, above] {$$} (fbk.west);
	
    	\draw [line] (fbk.east) -- node [midway, above] {$$} (fbegin.center);
	
    	\draw [line] (fbegin.center) -- node [midway, above] {$$} (dbegin.west);
	
\end{tikzpicture}
}
           \label{fig:bd-fbk1}
    \end{tabular}
	\\
    \begin{tabular}{c}
     \subfigure[]{
%
%
%
%
\input{tikz_common_styles}
%
%
\noindent
\begin{tikzpicture}[scale=1, auto, >=stealth']

	\node[] (zetaend) at (0,0) {};
	
	 \node[] (fend) at ($(zetaend.center) - (0.cm,1.5cm)$) {};
		 
    \node[block, minimum height = 1.cm, top color=blue!20, bottom color=blue!20] (Tstar) at ($(zetaend) + (3.65cm,0)$) 
    {$\dfrac{1}{\sigma} \, \MM{T}^{\dagger}(\omega)$};
    
    \node[block, minimum height = 1.cm, top color=blue!20, bottom color=blue!20] (T) at ($(Tstar) + (3.95cm,0cm)$) 
    {$\dfrac{1}{\sigma} \, \MM{T} (\omega)$};
    

     	\node[] (dbegin) at ($(T.east) + (2.cm,0)$) {};
	
     	\node[] (fbegin) at ($(dbegin.west) - (0.cm,1.5cm)$) {};
	
    	\draw [connector] (dbegin.west) -- node [midway, above] {$\tilde{\BB d} (y)$} (T.east);
    
    	\draw [connector] (T.west) -- node [midway, above] {$\tilde{\BB g} (y) \, = \; \tilde{\Bxi} (y)$} (Tstar.east);
	
    	\draw [line] (Tstar.west) -- node [midway, above] {$\tilde{\bzeta} (y) \, = \; \tilde{\BB d} (y)$} (zetaend);
	
    	\draw [line] (zetaend.east) -- node [midway, above] {$$} (fend.east);
	
    	\draw [line] (fend.east) -- node [midway, above] {$$} (fbegin.center);
	
    	\draw [line] (fbegin.center) -- node [midway, above] {$$} (dbegin.west);
	
\end{tikzpicture}
           \label{fig:bd-fbk2}}
    \end{tabular}
    \\
    \begin{tabular}{c}
     \subfigure[]{
%
%
%
%
\input{tikz_common_styles}
%
%
\noindent
\begin{tikzpicture}[scale=1, auto, >=stealth']

	\node[] (zetaend) at (0,0) {};
	
	 \node[] (fend) at ($(zetaend.center) - (0.cm,1.5cm)$) {};
		 
    \node[block, minimum height = 1.cm, top color=blue!20, bottom color=blue!20] (T) at ($(zetaend) + (2.5cm,0)$) 
    {$\dfrac{1}{\sigma} \, \MM{T}(\omega)$};
        
    \node[block, minimum height = 1.cm, top color=blue!20, bottom color=blue!20] (Tstar) at ($(fend) + (2.5cm,0)$) 
    {$\dfrac{1}{\sigma} \, \MM{T}^{\dagger} (\omega)$};

     	\node[] (dbegin) at ($(T.center) + (2.5cm,0)$) {};
	
     	\node[] (fbegin) at ($(dbegin.west) - (0.cm,1.5cm)$) {};
	
    	\draw [connector] (dbegin.west) -- node [midway, above] {$\tilde{\BB d} (y)$} (T.east);
    
    	\draw [line] (T.west) -- node [midway, above] {$\tilde{\Bxi} (y)$} (zetaend);
	
    	\draw [line] (zetaend.east) -- node [midway, above] {$$} (fend.east);
	
    	\draw [connector] (fend.east) -- node [midway, above] {$\tilde{\BB g} (y)$} (Tstar.west);
	
    	\draw [line] (Tstar.east) -- node [midway, above] {$\tilde{\bzeta} (y)$} (fbegin.center);
	
    	\draw [line] (fbegin.center) -- node [midway, above] {$$} (dbegin.west);
	
\end{tikzpicture}
           \label{fig:bd-fbk3}}
    \end{tabular}
\end{tabular}
  \caption{Through a sequence of transformations, the cascade connection of the operators $\MM{T}^{\dagger}(\omega)$ and $\MM T(\omega)$ shown in~\fref{fig:cascade} is cast as a feedback interconnection of $(1/\sigma) \, \MM{T}^{\dagger}(\omega)$ and $(1/\sigma) \, \MM T(\omega)$~\cite[Theorem 1]{boyd1989}.}
  	\label{fig:cas2feedback} 
	\vspace*{-0.5cm}
\end{figure}

	\vspace*{-2ex}
  In this section, we first summarize the standard procedure for computing the singular value decomposition of the frequency response operator $\MM T(\omega)$. This approach utilizes a cascade connection of $\MM T^{\dagger} (\omega)$ and $\MM T (\omega)$, shown in Figure~\ref{fig:cascade}, and it relies on computing inverses to determine the resolvent operator and its adjoint. Since it can suffer from ill-conditioning, we employ an alternative method that avoids inversion~\cite[Theorem 1]{boyd1989}. This method extends the standard reflection technique~\cite{lanczos1997linear,GolKahSIAM1965,AurTreSIAM2017} to frequency response analysis and exploits feedback interconnection, shown in Figure~\ref{fig:cas2feedback}, to avoid numerical errors and guard against ill-conditioning. We close the section with a discussion of numerical schemes that are utilized in this work.

The frequency response operator
	$
	\MM{T}(\omega)
	=
	\MM{C}(\omega)
	\MM A ^{-1}(\omega)
	\MM{B}(\omega)$
in~\eqref{eq:mot1} is described by
\begin{subequations}\label{eq:ops}
  \begin{equation}\label{eq:regularOp}
	 \Bxi(y)
	 \; = \;
	 [ \MM{T} \BB{d} (\cdot) ] (y)
	 ~~ \Leftrightarrow ~~
	\left\{
	\begin{aligned}
      \tc{black}{
      [ \MM{A} \, \bphi ( \cdot ) ] (y)
      }
      \;&=\;
      [ \MM{B} \,\BB{d} (\cdot) ] (y),
      &
      \\
  \Bxi(y)  \;&=\;  [ \MM{C}\,\bphi (\cdot) ] (y) ,&\\
[\MM{L}_a \, \bphi(\cdot)](a)  \;&=\;  [\MM{L}_b \, \bphi(\cdot)](b)  \; = \;  0, &
\end{aligned}
	\right.
\end{equation}
and the adjoint operator
	$
	\MM{T}^{\dagger}(\omega)
	=
	\MM{B}^\dagger(\omega)
	\MM{A}^{-\dagger}(\omega)
	\MM{C}^\dagger(\omega)
	$
is determined by
	\begin{equation}
	\label{eq:adjointOp}
	\bzeta(y)
	\; = \;
	[ \MM{T}^{\dagger} \BB{g} ( \cdot ) ] (y)
	~~ \Leftrightarrow ~~
	\left\{
	\begin{aligned}
     \tc{black}{[ \MM{A}^{\dagger} \bpsi ( \cdot ) ] (y)} \;&=\;  [ \MM{C}^\dagger \BB{g} ( \cdot ) ] (y),&\\
  \bzeta(y)  \;&=\;   [ \MM{B}^{\dagger} \bpsi ( \cdot ) ] (y),&\\
[\MM{L}_a^\dagger \bpsi (\cdot)](a)  \;&=\;  [\MM{L}_b^\dagger \bpsi(\cdot)](b)  \; = \;  0, &
\end{aligned} \right.
\end{equation}
\end{subequations}
where we suppress the dependence on $\omega$ for notational convenience. The adjoint operators are defined as~\cite{renardy2006introduction},
\begin{subequations}\label{eq:3.9}
\begin{align}
  \left<\bpsi,\MM{A}\bphi\right>  \;&=\;  \left<\MM{A}^\dagger \bpsi,\bphi\right>,\label{eq:3.9a}\\
  \left<\bpsi,\MM{B}\BB{d}\right>  \;&=\;  \left<\MM{B}^\dagger \bpsi,\BB{d}\right>,\label{eq:3.9b}\\
  \left<\BB{g},\MM{C}\bphi\right>  \;&=\;  \left<\MM{C}^\dagger\BB{g},\bphi\right>, \label{eq:3.9c}
\end{align}
\end{subequations}
where the boundary conditions on $\MM{L}_a^{\dagger}$ and $\MM{L}_b^{\dagger}$ in~\eqref{eq:adjointOp} are selected to ensure that~\eqref{eq:3.9a} holds. The analytical approach to computing the adjoint operators typically involves integration by parts whereas the numerical approach utilizes appropriate integration weights to make sure that the discrete approximation of the inner products in~\eqref{eq:3.9} holds true.

In~\cite{liejovJCP13}, the adjoints and the corresponding boundary conditions were evaluated analytically for arbitrary block matrix operators using the procedure described in~\cite[Section 5]{renardy2006introduction}. We note that a similar procedure as in~\cite{liejovJCP13} is also used in the current Chebfun system to compute the formal adjoint of a linear differential operator~\cite{driscoll2014chebfun}. While the method for determining formal adjoints described in~\cite[Section 5]{renardy2006introduction} and~\cite{liejovJCP13} can be also utilized for systems in the descriptor form, determination of the adjoint boundary conditions requires additional attention. For the linearized NS equations described in \S~\ref{sec:MotEx3}, the method developed in~\cite[Section 5]{renardy2006introduction} yields smaller number of boundary conditions than necessary to have a well-posed adjoint system. In \S~\ref{sec:bcs}, we describe how this challenge can be overcome by utilizing the governing equations to impose additional boundary conditions in order to make the adjoint system well-posed.  

The eigenvalue decomposition of the composite operator $\MM{T}^{\dagger}(\omega)\MM{T}(\omega)$, whose block diagram is shown in \fref{fig:cascade}, can be used to obtain  squares of the singular values. Detailed equations representing the composite operator can be found in~\cite{liejovJCP13}. Since the composite operator involves inverses of both $\MM A$ and $\MM A^{\dagger}$, computations can be prone to ill-conditioning. In the next section, we show how to conduct SVD of the frequency response operator $\MM{T}(\omega)$ without having to compute any inverses.

	\vspace*{-4ex}
\subsection{The feedback interconnection}\label{sec:feedback}

  \vspace*{-2ex}
Singular values of the matrix $\mathbf A\in \mathbb C^{n\times n}$ are typically computed via the eigenvalue decomposition of the matrix $\mathbf A \mathbf A^{\dagger}$ (or $\mathbf A^{\dagger} \mathbf A$)~\cite{golub2012matrix}. Alternatively, they can be obtained from the eigenvalues of the matrix~\cite{lanczos1997linear,GolKahSIAM1965,AurTreSIAM2017},
\begin{align*}
   \left[  \begin{array}{cc}
    0 & \mathbf{A} \\
    \mathbf{A}^{\dagger} & 0
   \end{array}\right].
\end{align*}
This so-called reflection technique avoids floating-point errors associated with computing the composite matrix $\mathbf A \mathbf A^{\dagger}$~\cite{lanczos1997linear,GolKahSIAM1965,AurTreSIAM2017}. In most cases this error is not significant and both methods should yield similar results. Since the frequency response operator and its adjoint involve inverses of the operators $\MM{A}$ and $\MM{A}^{\dagger}$, for ill-conditioned problems errors associated with computing these inverses can become large~\cite{golub2012matrix}. In what follows, we employ a method that is inspired by the reflection technique and provide a reformulation that does not involve any inversions~\cite[Theorem 1]{boyd1989}. 

Through a sequence of transformations, the composite system shown in~\fref{fig:cascade} can be brought into the feedback interconnection shown in the block diagram in Figure~\ref{fig:bd-fbk3}. This representation requires realizations of the operators $(1/\sigma) \, \MM T (\omega)$ and $(1/\sigma) \, \MM T^{\dagger} (\omega)$ which are respectively determined by
 \begin{subequations}\label{eq:3.8}
  \begin{equation}\label{eq:3.8a}
     \tilde{\Bxi} (y)
	 \; = \;
	 [ \tfrac{1}{\sigma} \, \MM{T} \, \tilde{\BB{d}} (\cdot) ] (y)
	 ~~ \Leftrightarrow ~~
    \left\{
    \begin{aligned} {} [ \MM{A} \, \tilde{ \bphi } ( \cdot ) ] (y)
          \;&=\;
          [ \MM{B} \, \tilde{ \BB{d} } (\cdot) ] (y),
          &
          \\
      \tilde{ \Bxi } (y)  \;&=\;  [ \tfrac{1}{\sigma} \, \MM{C}\, \tilde{ \bphi } (\cdot) ] (y) ,&\\
    [\MM{L}_a \, \tilde{ \bphi } (\cdot)](a)  \;&=\;  [\MM{L}_b \, \tilde{ \bphi } (\cdot)](b)  \; = \;  0, &
    \end{aligned}
      \right.
    \end{equation}
 and
 \begin{equation}\label{eq:3.8b}
  \tilde{\bzeta} (y)
	\; = \;
	[ \tfrac{1}{\sigma} \, \MM{T}^{\dagger} \tilde{\BB{g}} ( \cdot ) ] (y)
	~~ \Leftrightarrow ~~
  \left\{
  \begin{aligned}{}[ \MM{A}^{\dagger} \tilde{ \bpsi } ( \cdot ) ] (y)
         \;&=\;  [ \MM{C}^\dagger \tilde{ \BB{g} } ( \cdot ) ] (y),&\\
    \tilde{ \bzeta } (y)  \;&=\;   [ \tfrac{1}{\sigma} \, \MM{B}^{\dagger} \tilde{ \bpsi } ( \cdot ) ] (y),&\\
  [\MM{L}_a^\dagger \tilde{ \bpsi } (\cdot)](a)  \;&=\;  [\MM{L}_b^\dagger \tilde{ \bpsi } (\cdot)](b)  \; = \;  0, &
  \end{aligned} \right .
  \end{equation}
 \end{subequations}
The block diagram in~Figure~\ref{fig:bd-fbk3} requires setting $\tilde{\BB{d}}(y) = \tilde{\bzeta} (y)$ and $\tilde{\BB{g}} (y) = \tilde{\Bxi} (y)$ in~\eqref{eq:3.8}, which yields
\begin{subequations}\label{eq:resSys}
\begin{align}
  [\MM A^{\dagger} \tilde{\bpsi} (\cdot)](y)  \;&=\; [ \tfrac{1}{\sigma} \, \MM C^{\dagger} \MM C\, \tilde{\bphi} (\cdot)] (y),\\
  [\MM A\, \tilde{\bphi} (\cdot)](y)  \;&=\; [ \tfrac{1}{\sigma} \, \MM B \MM B^{\dagger} \tilde{\bpsi} (\cdot)](y).
\end{align}
\end{subequations}
This system can be equivalently expressed as the generalized eigenvalue problem,
\begin{align}\label{eq:matrixEvalProb}
\left[
\begin{array}{cc}
  0 &\MM{B}\MM{B}^{\dagger} \\
   \MM{C}^{\dagger}\MM{C}&0
\end{array} \right] \left[ \begin{array}{c}
  \tilde{\bphi} \\
  \tilde{\bpsi}
\end{array} \right]
\;=\;
\gamma \left[\begin{array}{cc}
  \MM{A} & 0\\
  0 & \MM{A}^{\dagger}
\end{array}\right]
\left[ \begin{array}{c}
  \tilde{\bphi} \\
  \tilde{\bpsi}
\end{array} \right],
\end{align}
where we suppress the dependence on the spatial variable $y$ for brevity. Eigenvalues resulting from this approach determine the singular values in pairs of opposite signs, i.e., $\gamma = \pm \sigma$. 

This approach offers two advantages relative to the computation of the eigenvalues using the composite operator $\MM{T}^{\dagger} (\omega)\MM{T}(\omega)$. First, it allows simultaneous computation of both the right and the left singular functions, i.e., $\tilde{\bphi} (y)$ and $\tilde{\bpsi} (y)$. The second and more important advantage is that \tcb{the use of the QZ algorithm} does not require computation of inverses~\cite{MolSteSIAM1973}. This feature avoids a potential issue of ill-conditioning and allows application to systems in the descriptor form, thereby avoiding the need for determining the evolution form representation. \tcb{However, these advantages come at the cost of having to solve a generalized eigenvalue problem that is twice the size compared to size resulting from the cascade interconnection in~\fref{fig:cascade}.}

In most cases, it is of interest to compute only the few largest \tcg{finite} singular values \tcb{and standard subspace-iteration techniques~\cite{saad1981krylov} can be used to accomplish this objective. We utilize well-conditioned spectral methods~\cite{OlvTowSIAM2013,DuSIAM2016} to obtain finite-dimensional approximations of the operators in the generalized eigenvalue problem~\eqref{eq:matrixEvalProb}. These methods typically lead to banded matrices which favorably reflects on sparsity of the discretized operators in~\eqref{eq:matrixEvalProb}. The use of sparse solvers requires one of the two matrices in~\eqref{eq:matrixEvalProb} to be invertible, a requirement that typically holds for physical systems. When both matrices in~\eqref{eq:matrixEvalProb} are non-invertible, certain sparse QZ algorithms~\cite{FokkemaJacobi-DavidsonSparseQZ} can be used to avoid matrix inversions.}


\tcb{In~\cite{monokrousosJFM2010}, a method that avoids matrix inversion was developed by recasting computation of the resolvent norm as an optimization problem which is solved using a time-stepping iterative procedure in conjunction with direct numerical simulations. In contrast, our method utilizes a direct approach to compute the resolvent norm, and is also closely related to computation of quick estimates for the $\MM H_{\infty}$ norm~\cite{boyd1989,BRUINSMA1990287}, i.e., the smallest upper bound on the largest singular value of the frequency response operator across temporal frequencies.}

When $\MM A(\omega) = \mri \omega \MM E - \MM F$, we next describe how the procedure of this section can be utilized to compute the $\MM H_\infty$ norm. For stable linear dynamical systems, this quantity determines the $L_2$-induced gain (i.e., the worst-case amplification of finite energy disturbances) and it has an appealing robustness interpretation~\cite[Section 4.10.2]{skogestad2007multivariable} that is closely related to the notion of pseudo-spectra of linear operators~\cite{treemb05}. 

	\vspace*{-4ex}
  \subsubsection*{Computation of the $\MM H_\infty$ norm}\label{sec:HinfNorm}
    
  	\vspace*{-2ex}
 The peak of the largest singular value of the frequency response operator ${\MM T} (\omega)$ over all temporal frequencies $\omega \in \mathbb R$ determines the $\MM H_{\infty}$ norm of a stable linear time-invariant system,
\begin{equation}\label{eq:HinfNorm}
  \| \, \MM T \, \|_{\infty} \;\DefinedAs\; \sup\limits_{\omega}  ~\sigma_{\max}(\MM T (\omega)).
\end{equation}
When $\MM A(\omega) = \mri \omega I - \MM F$, the $\MM H_{\infty}$ norm can be computed to a desired accuracy using the purely imaginary eigenvalues of the Hamiltonian operator~\cite{boyd1989,BRUINSMA1990287}, 
\begin{equation}\label{eq:Hamiltonian}
  \MM M_\gamma 
  \;=\; 
  \left[
  \begin{array}{cc}
   \MM F & \tfrac{1}{\gamma} \, \MM B \MM B^\dagger
   \\[0.15cm]
   - \tfrac{1}{\gamma} \, {\MM C}^\dagger {\MM C} & -\MM F^\dagger 
  \end{array}\right]. 
\end{equation}
For a given $\omega = \omega_0$, the formulation based on a feedback interconnection~\eqref{eq:matrixEvalProb} implies that $\gamma$ is a singular value of $\MM T (\omega_0)$. The expression for $\MM M_\gamma $ given by~\eqref{eq:Hamiltonian} can be obtained by rearranging~\eqref{eq:matrixEvalProb}, and a selected value of $\gamma = \gamma_1$ is a singular value of $\MM T(\omega)$ if and only if $\MM M_{\gamma_1}$ has at least one purely imaginary eigenvalue~\cite[Theorem 2]{boyd1989}. In this case, $\gamma_1$ provides a lower bound on $\| \, \MM T \, \|_{\infty}$ and the value of $\gamma_1$ can be updated using either the bi-section algorithm~\cite{boyd1989} or the method provided in~\cite{BRUINSMA1990287} to compute the $H_\infty$ norm to a desired~accuracy.

This procedure can be also extended to the problems with $\MM A(\omega) = \mri \omega \MM E - \MM F$; e.g., see~\cite{BennerDescriptor2012}. The algorithm involves calculation that identifies the existence of purely imaginary eigenvalues for a generalized eigenvalue problem with operators $\left(\MM M_\gamma,\MM N_\gamma\right)$, where, 
\begin{equation}\label{eq:HamiltonianD}
  \MM N_\gamma \;=\; \left[\begin{array}{cc}
   \MM E & 0\\[0.15cm]
   0 & \MM E^\dagger 
  \end{array}\right].
  \non
\end{equation}
	
	\vspace*{-6ex}
\subsection{Numerical approximation of spatial differential operators}
 	\label{sec:numMeth1discretization}

 	\vspace*{-2ex}
Solving two-point boundary value problems via spectral methods requires expressing the variable of interest in a global basis of orthogonal functions, e.g., the Chebyshev polynomials. For example, in reaction-diffusion equation~\eqref{eq:rnd} the variable $\phi(y,t)$ can be expressed as
\begin{align*}
  \phi (y,t)  \;=\; \sideset{}{'}\sum_{i \, = \, 0}^{\infty} a_i (t) T_i(y),
\end{align*}
where $T_n(y)$ is the $n$th Chebyshev polynomial of the first kind, $a_n (t)$ is the $n$th spectral coefficient, and the prime denotes a summation with the first term halved.

		\vspace*{-4ex}
\subsubsection{Implementation using Chebfun}\label{sec:chebfun}

	\vspace*{-2ex}
Chebfun is an open-source software for spectral methods that provides various standard discretizations~\cite{driscoll2014chebfun}.
We implement the feedback interconnection shown in~\fref{fig:cas2feedback} using Chebfun in {\sf Matlab}~\cite{driscoll2014chebfun,driscoll2010automatic} and explore the utility of different discretization schemes that Chebfun offers. As a representative of an ill-conditioned discretization scheme, we use Chebfun's spectral collocation routine which utilizes Chebyshev polynomials of the second kind as basis functions and goes under the name {\sf chebcolloc2}. Chebfun also provides a well-conditioned scheme, {\sf ultraS}, which expresses the $k$th derivative of a function in terms of a series of ultraspherical polynomials~\cite{OlvTowSIAM2013}. We develop a function that takes the operators $\MM A$, $\MM B$, and $\MM C$ in~\eqref{eq:mot1} as inputs in the Chebfun syntax, and produces the singular values and the corresponding singular functions as outputs. For systems with $\MM A(\omega) = \mri \omega \MM E - \MM F$ and nonsingular $\MM E$, we also provide a function that computes the $\MM H_\infty$ norm and returns the frequency at which $\sigma_{\max} (\omega)$ peaks using the algorithm developed in~\cite{BRUINSMA1990287}. All routines that utilize Chebfun are restricted to systems in the evolution form. 

	\vspace*{-4ex}
\subsubsection{Implementation using spectral integration suite}\label{sec:sis}

  \vspace*{-2ex}
We develop a spectral integration suite that implements the feedback interconnection whose block diagram is shown in~\fref{fig:bd-fbk3}. The suite is based on the methods reported in~\cite{GreSIAM91,DuSIAM2016} with minor modifications that facilitate application to a broad class of infinite-dimensional problems and results in simple discretization matrices in \textsf{Matlab} and {\sf C++}. As discussed in~\sref{sec:MotEx3}, in contrast to conventional spectral methods, the spectral integration method is attractive because it does not require a staggered grid to deal with systems in the descriptor form. In the remainder of this section, we provide a brief summary of our implementation of the Chebyshev spectral integration method and relegate details to supplementary material. 

In the spectral integration method, the highest derivative is expressed in the basis of Chebyshev polynomials (in our case, of the first kind) and expressions for lower derivatives are determined by integrating higher derivatives. For the reaction-diffusion equation~\eqref{eq:rnd}, the second derivative of $\phi(y)$ is expressed as
\begin{subequations}\label{eq:specInt}
\begin{equation}
  \MD^2\phi (y) \;=\; \sideset{}{'}\sum_{i \, = \, 0}^{\infty} \phi_i^{(2)} T_i(y)\; \AsDefined \;
  \HH t_y^T \BB \Phi^{(2)},\label{eq:specInt1}
\end{equation} 
where $\BB \Phi^{(2)} = [\,\phi^{(2)}_0\;\; \phi^{(2)}_1\;\; \phi^{(2)}_2 \;\; \cdots \;\; ]^T$ is the infinite vector of spectral coefficients and $\HH t_y$ is the vector of Chebyshev polynomials of the first kind $T_i(y)$,
	$
  	\HH t_y^T \DefinedAs \left[\,\tfrac{1}{2} T_0(y) \;\; T_1(y) \;\; T_2(y) \;\;  \cdots \;\;\; \right].
	$
Subsequent indefinite integration of~\eqref{eq:specInt1} yields
\begin{IEEEeqnarray}{rclrl}
  	\MD \phi (y) 
  	&~\, = ~\,& 
 	 \displaystyle \sideset{}{'}\sum_{i \, = \, 0}^{N} \phi_i^{(1)} T_i(y) \,+\, c_1
  	&~\, \AsDefined ~\,&
  	\HH t_y^T \BB \Phi^{(1)}\,+\, c_1,
	\label{eq:specInt2}\\
	\phi (y) 
	&~\, = ~\,& 
	\displaystyle \sideset{}{'}\sum_{i \, = \, 0}^{N} \phi_i^{(0)} T_i(y) \,+\, c_1 y \,+\, \tilde c_0 
	&~\, \AsDefined ~\,& 
	\HH t_y^T \BB \Phi^{(0)}\,+\, c_1 y \,+\, \tilde c_0,\label{eq:specInt3}
\end{IEEEeqnarray}
\end{subequations}
where $\tilde c_0$ and $c_1$ are constants of integration. The spectral coefficients of $\BB \Phi^{(1)}$ and $\BB \Phi^{(0)}$ are related to the spectral coefficients of $\BB \Phi^{(2)}$ as
\begin{IEEEeqnarray}{c}
  \BB \Phi^{(1)} \;=\; \mathbf {Q} \, \BB \Phi^{(2)}, \quad \BB \Phi^{(0)} \;=\; {\mathbf Q}^2 \BB \Phi^{(2)}, 
\end{IEEEeqnarray}
where ${\mathbf Q}$ is given by,
\begin{equation}\label{eq:Q}
  {\mathbf Q} \;\DefinedAs\; \left[\begin{array}{ccccccc}
     0& \tfrac{1}{2} & 0 & \cdots\\
    \tfrac{1}{2} & 0 & -\tfrac{1}{2} & 0 &\cdots \\
    0 & \tfrac{1}{4} & 0 & -\tfrac{1}{4} & 0 &\cdots \\
    0 & 0 & \tfrac{1}{6} & 0 & -\tfrac{1}{6} & 0 &\cdots \\
    \vdots & \vdots &  & \ddots & \ddots & \ddots & \\
  \end{array}\right].
\end{equation}
The first row of the integration operator ${\mathbf Q}$ we use in~\eqref{eq:Q} is different from what is used in~\cite[Section~4]{DuSIAM2016} and~\cite[Eq.~(12)]{GreSIAM91}, and its derivation is provided in supplementary material. In contrast to~\cite{GreSIAM91,DuSIAM2016} where the first row of $\mathbf Q$ is full, our representation for $\mathbf Q$ in~\eqref{eq:Q} is given by a banded tri-diagonal matrix. 

From the above, we can express $\phi (y)$, $\MD \phi (y)$, and $\MD^2 \phi (y)$ and  as
\begin{subequations}\label{eq:rndus}
  \begin{align}
  \phi(y) 
  \;&=\; 
   \HH t_y^T ( {\mathbf Q}^2 \HH \Phi^{(2)} \, + \, \R_2 \HH c ),
   \label{eq:rndu}
	\\
	\MD\, \phi (y) 
  \;&=\; 
	\HH t_y^T ( {\mathbf Q}^1 \HH \Phi^{(2)} \,+ \, \R_1 \HH c ),
	\label{eq:rndDu}
	\\
	\MD^2 \phi (y) 
  \;&=\; 
	\HH t_y^T ( {\mathbf Q}^0 \HH \Phi^{(2)} \,+ \, \R_0 \HH c ),
	\label{eq:rndD2u}
\end{align} 
\end{subequations}
where $\R_i$ are matrices that account for the constants of integration in a basis of Chebyshev polynomials, $\BB c \DefinedAs  [ \, c_0 \;\; c_1 \, ]^T$, and $c_0 = 2 \tilde c_0$.

The feedback interconnection used to compute the frequency response of~\eqref{eq:rnd} is given by (see~\eqref{eq:matrixEvalProb}),
\begin{subequations}\label{eq:feedbackRnd}
\begin{align}
  \left[\begin{array}{cc}
    0 & I\\
    I & 0
  \end{array}\right]\left[\begin{array}{c}
    \phi(y)\\
    \psi(y)
  \end{array}\right]\;&=\;\gamma \,\left[\begin{array}{ccc}
    (\mri \omega + \epsilon^2) I \,-\, \MD^2 & & 0\\
    0 && (-\mri \omega + \epsilon^2) I \,-\, \MD^2
  \end{array}\right] \left[\begin{array}{c}
    \phi(y)\\
    \psi(y)
  \end{array}\right],\label{eq:feedbackRndEq}\\
  \left[\begin{array}{cc}
    {\MM L}(+1,\MD) & 0\\
    {\MM L}(-1,\MD)& 0\\
    0 & {\MM L}(+1,\MD)\\
    0 & {\MM L}(-1,\MD)
  \end{array}\right]\left[\begin{array}{c}
    \phi(y)\\
    \psi(y)
  \end{array}\right] \;&=\; 0,\label{eq:feedbackRndBc}
\end{align}
\end{subequations}
where ${\MM L}(a,L)$ evaluates the action of the linear operator $L$ on a variable at a point $y = a$. In particular,~\eqref{eq:feedbackRndBc} specifies homogeneous Neumann boundary conditions at $y = \pm 1$. 

\begin{subequations}\label{eq:rnd_fr}
  For the reaction-diffusion equation, the infinite-dimensional representation of the system shown in Figure~\ref{fig:cas2feedback} is obtained by combining~\eqref{eq:rndus} with~\eqref{eq:feedbackRndEq} and equating terms that correspond to the same basis functions,
  \begin{IEEEeqnarray}{rCl}\label{eq:rnd_fr_eq}
    \underbrace{\left[\begin{array}{cccc}
      \0 & \0 & {\HH Q}^2 & \R_2\\
      {\HH Q}^2 & \R_2 & \0 & \0
    \end{array}\right]}_{\hat{\HH E}}\,\underbrace{\left[\begin{array}{c}
      \BB \Phi^{(2)}\\
      \HH c^{\phi}\\
      \BB \Psi^{(2)}\\
      \HH c^{\psi}
    \end{array}\right]}_{\hat{\HH v}}
    ~&=&~ 
    \nonumber 
    \\[-0.25cm]
    \IEEEeqnarraymulticol{3}{c}{
      \gamma 
    \underbrace{\left[\begin{array}{cccccc}
      (\mri \omega + \epsilon^2){\HH Q}^2 - \I && (\mri \omega + \epsilon^2) \R_2 - \R_0 & \0 && \0\\
      \0 && \0 & (-\mri \omega + \epsilon^2) {\HH Q}^2 - \I && (-\mri \omega + \epsilon^2) \R_2 - \R_0
    \end{array}\right]}_{\hat{\HH F}} \underbrace{\left[\begin{array}{c}
      \BB \Phi^{(2)}\\
      \HH c^{\phi}\\
      \BB \Psi^{(2)}\\
      \HH c^{\psi}
    \end{array}\right]}_{\hat{\HH v}}},
  \end{IEEEeqnarray} 
    
Similarly, substitution of~\eqref{eq:rndus} to~\eqref{eq:feedbackRndBc} yields the representation of boundary conditions,
\begin{IEEEeqnarray}{rCl}\label{eq:rnd_fr_bc}
 \underbrace{ \left[\begin{array}{cccc}
    \HH t_{+1}^T\,{\HH Q} & \HH t^T_{+1}\,\R_1 & \0 & \0\\[0.15cm]
    \HH t_{-1}^T\,{\HH Q} & \HH t^T_{-1}\,\R_1 & \0 & \0\\[0.15cm]
     \0 & \0& \HH t_{+1}^T\,{\HH Q} & \HH t^T_{+1}\,\R_1\\[0.15cm]
     \0 & \0&\HH t_{-1}^T\,{\HH Q} & \HH t^T_{-1}\,\R_1 \\
  \end{array}\right]}_{\hat{\HH M}}\left[\begin{array}{c}
    \BB \Phi^{(2)}\\
    \HH c^{\phi}\\
    \BB \Psi^{(2)}\\
    \HH c^{\psi}
  \end{array}\right] \;&=&~ \0.
\end{IEEEeqnarray}
\end{subequations}	
Thus, in the generalized eigenvalue problem~\eqref{eq:rnd_fr_eq} only the eigenfunctions that belong to the null-space of the operator in~\eqref{eq:rnd_fr_bc} are acceptable and the system of equations~\eqref{eq:rnd_fr} can be written as,
	\beq
	\ba{rcl}
	\hat{\HH E}\,\hat{\HH v} 
	~&=&~ 
	\gamma \,\hat{\HH F}\, \hat{\HH v},
	\\[0.1cm]
	\hat{\HH M }\,\hat{\HH v} 
	~&=&~ 
	\0.
	\ea
	\label{eq:rnd_fr_abstract}
	\eeq
The finite-dimensional approximation of~\eqref{eq:rnd_fr_abstract} is derived by utilizing a projection operator, 
\begin{equation}\label{eq:project}
  \hat{\HH P} \;=\; \left[\,\I_{N+1}\;\; \0 \,\right],
\end{equation}
where $ \hat{\HH P}$ has $N + 1$ rows and an infinite number of columns. The projection operator~\eqref{eq:project} is applied to the spectral coefficients of the regular and adjoint variables and not the constants of integration. \mbox{We use matrices}
\beq
\hat{\HH R} \;=\; \left[\begin{array}{cccc}
  \hat{\HH P} & 0 & 0 & 0\\
  0 & \I_2 & 0 & 0\\
  0 & 0 & \hat{\HH P} & 0\\
  0 & 0 & 0 & \I_2
\end{array}\right] , 
	~~
\hat {\HH P}_2 \;=\; \left[\begin{array}{cc}
  \hat{\HH P} & \0\\
  \0 & \hat{\HH P}
\end{array}\right],
\eeq
to reduce~\eqref{eq:rnd_fr_abstract} to
\begin{subequations}\label{eq:rnd_fr_abstract_finite}
\begin{align}
\HH E\,\HH v 
\;&=\;
\gamma \,\HH F\, \HH v,\label{eq:rnd_fr_abstract_finite_a}
\\[0.cm]
\HH M \,\HH v 
\;&=\;
\0,\label{eq:rnd_fr_abstract_finite_b}
\end{align}
\end{subequations}
where,
\begin{align*}
  \HH E \; \DefinedAs \; \hat{\HH P}_2 \hat{\HH E} \, \hat{\HH R}^{T},
  ~~
  \HH F \; \DefinedAs \; \hat{\HH P}_2 \hat{\HH F}\, \hat{\HH R}^{T},
  ~~
  \HH M \; \DefinedAs \; \hat{\HH M} \,\hat{\HH R}^{T},
  ~~
  \HH v \; \DefinedAs \; \hat{\HH R}\,\hat{\HH v}.
\end{align*}

The SVD of the fat full-row-rank matrix $\HH M $ in~\eqref{eq:rnd_fr_abstract_finite_b} can be used to parameterize its null-space and obtain the eigenfunctions that satisfy the boundary conditions~\eqref{eq:feedbackRndBc}~\cite{jovbamSCL06} (see supplementary material for details), 
	\begin{equation}
  \HH M\,\HH v 
  \; = \; 
  \HH U \BB \Sigma \HH V^\dagger \HH v 
  \; = \; 
  \HH U  \left[\begin{array}{cc} \BB \Sigma_1 & \0 \end{array} \right] \left[\begin{array}{c}\HH V_{1}^\dagger \\[0.3em] \HH V_{2}^\dagger \end{array}\right] \HH v 
  \; = \; 
  0.\label{eq:Msvd}
\end{equation}
Thus, $\HH v \DefinedAs \HH V_{2} \HH u$ parametrizes the null-space of the matrix $\HH M$~\cite{FourSubspaces} and satisfies Eq.~\eqref{eq:rnd_fr_abstract_finite_b}.  
Substituting this expression for $\HH v$ in~\eqref{eq:rnd_fr_abstract_finite_a} yields the finite-dimensional generalized eigenvalue problem,
\begin{align}
  (\HH E \HH V_{\! 2})\,\HH u \;=\; \gamma \,(\HH F \HH V_{\! 2})\, \HH u,
\end{align}
which can be used to compute the singular values as $\gamma = \pm \sigma$ and $\HH u$.
	
Finite-dimensional approximations of more complex systems, e.g., the linearized NS equations~\eqref{eq:lns} and the equations governing channel flow of a viscoelastic fluid~\eqref{eq:2Dvisco}, are derived using a similar procedure. Additional care is required to account for spatially varying coefficients and for the presence of a static-in-time constraint that arises from the continuity equation. An in-depth discussion of our implementation of spectral integration in both {\sf C++} and {\sf Matlab} is provided in the supplementary material. Finally, we solve a generalized eigenvalue problem resulting from the finite-dimensional approximation to~\eqref{eq:matrixEvalProb} using the sparse eigenvalue solver, {\sf eigs} in {\sf Matlab}, and LAPACK's {\sf zggev} routine in {\sf C++}. 

As discussed in~\sref{sec:feedback}, the feedback interconnection in~\fref{fig:bd-fbk3} can be used for systems in the descriptor form and the spectral integration method does not require a staggered grid when pressure is retained in the governing equations. We next describe how we handle pressure boundary conditions in the spectral integration method for channel flows of incompressible Newtonian and viscoelastic fluids. 

\vspace*{-4ex}  
\subsection{Boundary conditions for the linearized NS equations in the descriptor form}\label{sec:bcs}

	\vspace*{-2ex}
\subsubsection{Boundary conditions for the frequency response operator}\label{sec:bcsReg}

\vspace*{-2ex}
For the linearized NS equations in the descriptor form, the boundary conditions on pressure are unknown and it is necessary to impose additional constraints to guarantee well-posedness. These additional boundary conditions do not need to be imposed on pressure fluctuations~\cite{AurTreSIAM2017}. In particular, the no-slip and no-penetration conditions at the walls, $\BB v (\pm 1) = 0$, can be used in conjunction with continuity equation~\eqref{eq:lnsb} (i.e., $\mri k_x u(y) + \DD v(y) + \mri k_z w(y) = 0$ after the Fourier transform in the wall-parallel directions has been utilized) to obtain two additional constraints,  
	$
	[\DD v(\cdot)](\pm 1)  =  0.
	$
Thus, the velocity fluctuations in the descriptor formulation of the NS equations have to satisfy eight boundary conditions,
\begin{align} 
	\label{eq:lnsBcFin}
    u(\pm 1) \, = \, v(\pm 1) \, = \, w(\pm 1) \, = \, [\DD v(\cdot)](\pm 1) \, = \, 0.
  \end{align}
The number of integration constants has to be equal to the number of (linearly independent) constraints for the spectral integration method to ensure well-posed numerical implementation~\cite{DuSIAM2016}. Since $\DD^2 \BB v$ and $\DD p$ appear in~\eqref{eq:lns}, expressing them in terms of Chebyshev polynomials and integrating would give one integration constant less than the number of boundary conditions. A well-posed formulation can be obtained by expressing the second derivative of the pressure in a basis of Chebyshev polynomials,
  \begin{equation}
    \DD^2 p (y) \;=\; \sideset{}{'}\sum_{i \, = \, 0}^{N} p_i^{(2)} T_i(y). \label{eq:specIntp}
  \end{equation} 
Subsequent integration (as in~\eqref{eq:specInt2} and~\eqref{eq:specInt3}) yields two additional integration constants which can be used to account for $[\DD v(\cdot)](\pm 1)  =  0$. Such a treatment for pressure is not uncommon in numerical approximations of the linearized NS equations; for example, two homogeneous Neumann boundary conditions on pressure have been used for modal analysis of the formulation in primitive variables~\cite{pressBCincom,OrszagBook,KHORRAMI1989206}. 
 
While conventional spectral methods (e.g., the Chebyshev-tau and collocation methods) require different numbers of basis functions for pressure and velocity fluctuations (i.e., a staggered grid) to avoid spurious modes~\cite{CanHusQuaTho2012,boy2001}, we express velocity and pressure using an equal number of basis functions, i.e., $N +1$. Moreover, the additional Neumann boundary conditions on wall-normal velocity fluctuations simply result from imposing the no-slip and no-penetration conditions at the walls, $\BB v(\pm 1)$, on the continuity equation~\eqref{eq:lnsb}. The same process of deriving linearly independent boundary conditions to make a spectral collocation method well-posed was previously used in pipe flow~\cite{khorrami1991chebyshev}. However, in contrast to the spectral integration method, the spectral collocation technique still requires a staggered grid~\cite{khorrami1991chebyshev}. 

In summary, we augment the linearized NS equations~\eqref{eq:lns} with boundary conditions~\eqref{eq:lnsBcFin}. In~\sref{sec:app3a}, we demonstrate that these boundary conditions produce the correct eigenvalues for the formulation in primitive variables (i.e., the descriptor form of the linearized NS equations) without a staggered grid. 

	\vspace*{-4ex}
\subsubsection{Boundary conditions for the adjoint system}\label{sec:bcsAd}

	\vspace*{-2ex}
For the NS equations linearized around the base flow $(U(y),0,0)$, application of the Fourier transform in $t$, $x$, and $z$ on~\eqref{eq:lns} yields the operators $\MM A$, $\MM B$, and $\MM C$ in~\eqref{eq:regularOp},
\begin{equation}\label{eq:LNSReg}
	\ba{rcl}
	\MM A
	& = &
  \left[\begin{array}{cccc}
    \mri ( \omega\, +  k_x U) - \tfrac{\Delta}{Re} &U'(y) & 0 & \mri k_x\\
    0 & \mri ( \omega\, +  k_x U) - \tfrac{\Delta}{Re} &  0 & {\mathrm D} \\
    0 & 0 & \mri ( \omega\, +  k_x U) - \tfrac{\Delta}{Re} & \mri k_z\\
    \mri k_x & {\mathrm  D} & \mri k_z & 0
  \end{array}\right],
  \\[0.35cm]
  \MM B
	& = &
	 \left[\begin{array}{cccc}
    I & 0 & 0 \\
    0 & I &  0 \\
    0 & 0 & I \\
    0 & 0 & 0
  \end{array}\right],
  ~~
  \MM C
	\; = \;
	 \left[\begin{array}{cccc}
    I & 0 & 0 & 0 \\
    0 & I &  0 & 0 \\
    0 & 0 & I & 0  
    \end{array}\right],
  \ea
\end{equation}
where $\Delta \DefinedAs \DD^2 - (k_x^2 + k_z^2)I$. The operators $\MM A$ and $\MM C$ act on the vector of flow fluctuations in primitive variables, i.e., $\bphi = [\,u ~ v ~ w ~ p\,]^T$ in~\eqref{eq:regularOp}; the operator $\MM B$ acts on the vector of forcing fluctuations, $\BB{d} = [\, d_x ~ d_y ~ d_z \,]^T$; and the output is determined by the velocity fluctuation vector, $\Bxi = \BB{v} = [\,u ~ v ~ w \,]^T$. Following~\cite[Section 5]{renardy2006introduction}, we obtain the adjoint operators $\MM A^\dagger$, $\MM B^\dagger$, and $\MM C^\dagger$ in~\eqref{eq:adjointOp}, 
	\begin{equation}
	\label{eq:LNSAd}
	\ba{rcl}
	\MM A^\dagger
	& = &
  \left[\begin{array}{cccc}
    - \mri ( \omega\, +  k_x U) - \tfrac{\Delta}{Re} & 0 & 0 & - \mri k_x\\
    U'(y) & - \mri ( \omega\, +  k_x U) - \tfrac{\Delta}{Re} &  0 & - {\mathrm D} \\
    0 & 0 & - \mri ( \omega\, +  k_x U) - \tfrac{\Delta}{Re} & - \mri k_z\\
    - \mri k_x & -{\mathrm D} & - \mri k_z & 0
  \end{array}\right],
  \\[0.35cm]
  \MM B^\dagger
	& = &
	\left[\begin{array}{cccc}
    I & 0 & 0 & 0 \\
    0 & I &  0 & 0 \\
    0 & 0 & I & 0  
    \end{array}\right],
  ~~
  \MM C^{\dagger}
	\; = \;
	\left[\begin{array}{cccc}
    I & 0 & 0 \\
    0 & I &  0 \\
    0 & 0 & I \\
    0 & 0 & 0
  \end{array}\right],
  \ea
\end{equation}
and show that the adjoint variables $\bpsi = [\,\hat u ~ \hat v ~ \hat w ~ \hat p\,]^T$ in~\eqref{eq:adjointOp} satisfy $\hat u (\pm 1) = \hat v (\pm 1) = \hat w (\pm 1) = 0$. Furthermore, evaluation of the last row in $\tc{black}{[ \MM{A}^{\dagger} \bpsi ( \cdot ) ] (y)} =  [ \MM{C}^\dagger \BB{g} ( \cdot ) ] (y)$ at the walls yields two additional boundary conditions $\DD\hat v (\pm 1) = 0$. Thus, for the linearized NS equations in the descriptor form we impose the following boundary conditions on the components of the vector $\bpsi = [\,\hat u ~ \hat v ~ \hat w ~ \hat p\,]^T$ in~\eqref{eq:adjointOp},
  \begin{align}\label{eq:lnsBcAdFin}
    \hat u(\pm 1) \, = \, \hat  v(\pm 1)  \, = \, \hat  w(\pm 1)  \, = \,  [\DD\hat  v(\cdot)](\pm 1)   \, = \,  0.
  \end{align}
 In~\sref{sec:app3a}, we demonstrate that the spectral integration method with boundary conditions~\eqref{eq:lnsBcFin} on the frequency response operator along with the adjoint boundary conditions~\eqref{eq:lnsBcAdFin} can be used to correctly compute the resolvent norm for the linearized NS equations in the descriptor form. We next show how this formulation can be extended to viscoelastic fluids. 
  
	\vspace*{-4ex}
\subsection{Frequency response analysis of 3D channel flow of a viscoelastic fluid}\label{sec:FreqResVisco}

	\vspace*{-2ex}
 The flow of a viscoelastic fluid in a channel is governed by equations \eqref{eq:2.1c} that account for the memory (time-dependent variation) of the stress in the fluid. As there are no boundary conditions on stress fluctuations, it is favorable to transform~\eqref{eq:2.1} in a manner that the stress is eliminated, and to retain as state variables quantities whose boundary conditions are known, i.e., the velocity and pressure fluctuations (as discussed in~\sref{sec:bcs}, velocity boundary conditions derived from the continuity equation account for pressure boundary conditions). After a spatio-temporal Fourier transform, the stress can be expressed in terms of the velocity~as,
 \begin{equation}\label{eq:tauTov}
   \BB \tau (y) \;=\; [\mathcal {V}\,\BB v(\cdot) ] (y),
   \end{equation}
The derivation of the operator $\MM V$ is described in Appendix~\ref{sec:appc} and two approaches to compute frequency responses, that rely on elimination of stress fluctuations, are discussed next.
	
		\vspace*{-4ex}
\subsubsection{The descriptor formulation with the stress eliminated}\label{sec:desStressElim}
	
	\vspace*{-2ex}
In this approach, we utilize~\eqref{eq:tauTov} to derive a system equivalent to \eqref{eq:2.1}. This yields a system of equations with state variables $\tilde{\bphi} = [\,u\;v\;w\; p\,]^T$ in~\eqref{eq:regularOp}. The operators $\MM A$, $\MM B$, and $\MM C$ in \eqref{eq:regularOp} for this system are given in Appendix \ref{sec:appCDes} and the boundary conditions are the same as that for linearized NS equations, i.e., \eqref{eq:lnsBcFin} and \eqref{eq:lnsBcAdFin}. In this paper, the ``descriptor form'' for viscoelastic fluids refers to the formulation in which stress fluctuations have been eliminated.

	\vspace*{-4ex}
\subsubsection{The evolution form model}\label{sec:evolForm}

	\vspace*{-2ex}
Once the stress fluctuations have been eliminated using the procedure described in~\S~\ref{sec:desStressElim}, the pressure can also be eliminated to bring the 3D viscoelastic system to a form where the state variables are given by $\tilde{\bphi} = [\,v\;\eta\,]^T$ with $\eta \DefinedAs \mri k_z u - \mri k_x w$~\cite{jovbamJFM05,schmid2012stability,jovARFM20}. This system is now in the evolution form where, apart from the boundary conditions, there are no additional constraints on state variables and the standard procedure described in~\cite[Section 5]{renardy2006introduction} can be used to determine the adjoint boundary conditions. The corresponding system representation~\eqref{eq:regularOp} is given in Appendix~\ref{sec:appCEvol}. In this paper, the evolution form for viscoelastic fluids refers to the formulation in which stress \tcg{and pressure} fluctuations have been eliminated.

	\vspace*{-3ex}
\section{Examples}\label{sec:apps}

\vspace*{-2ex}
In this section, we provide examples to demonstrate the merits and the effectiveness of the developed framework. For the reaction-diffusion equation, we show that the computations based on a feedback interconnection shown in Figure~\ref{fig:bd-fbk3} are insensitive to the operator-induced ill-conditioning discussed in~\sref{sec:MotEx1}. We next apply this feedback interconnection to the 2D viscoelastic system in the evolution form~\eqref{eq:2Dvisco} and show that our approach provides robust results over a much wider range of elasticities than the approach based on a cascade connection shown in Figure~\ref{fig:cascade}. Finally, we use the feedback interconnection in conjunction with the spectral integration method to compute frequency responses of systems in the descriptor form. 

	\vspace*{-3ex}
\subsection{Reaction-diffusion equation}\label{sec:apps1}

\begin{figure}
  \centering
  \includegraphics[scale = 0.25]{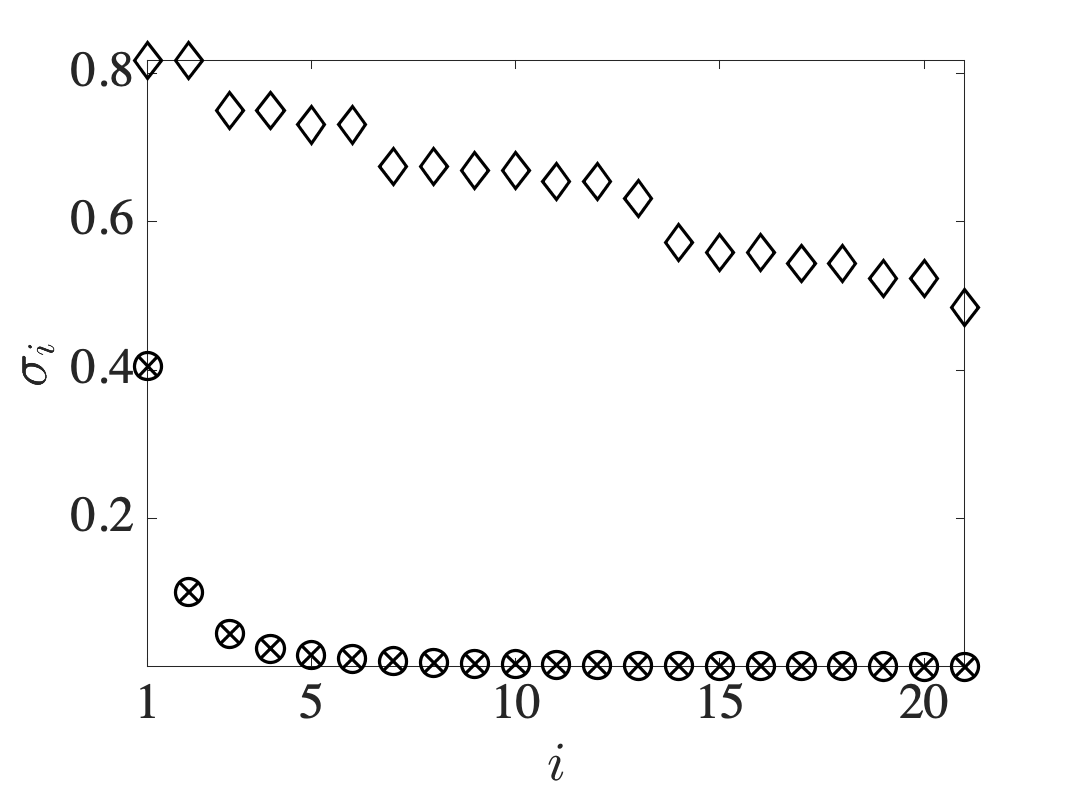}
    \caption{\label{fig:res1}Singular values of the frequency response operator for reaction-diffusion equation~\eqref{eq:rnd} with $\epsilon = 10^{-4}$ and $\omega = 0$ resulting from the use of Chebfun's spectral collocation scheme with $N = 64$. Symbols represent analytical solution ($\times$), and the computations based on the feedback interconnection shown in~\fref{fig:bd-fbk3} ($\circ$) and the cascade connection shown in Figure~\ref{fig:cascade} ($\Diamond$). The principal singular value (corresponding to $i = 0$) is not shown as its value is significantly larger than the remaining singular values.}
    \label{fig:res1b}
    \vspace*{-0.5cm}
  \end{figure}

	\vspace*{-2ex}
As demonstrated in \S~\ref{sec:MotEx1}, SVD of the operator $\MM T(\omega)\MM T^{\dagger}(\omega)$ is ill-conditioned for small values of $\epsilon$ in~\eqref{eq:rnd}. We revisit this example using the feedback interconnection shown in~Figure~\ref{fig:bd-fbk3}. \fref{fig:res1b} shows the first twenty singular values of the frequency response operator for reaction-diffusion equation~\eqref{eq:rnd} with $\epsilon = 10^{-4}$. While the values computed using the feedback connection (marked by circles) agree with the analytical solution (marked by crosses), the singular values resulting from the cascade connection (marked by diamonds) are erroneous. This mismatch arises from ill-conditioning of the operator $\MM T(\omega)\MM T^{\dagger}(\omega)$ and has nothing to do with the spatial discretization (increasing $N$ does not improve computations resulting from the cascade connection shown in Figure~\ref{fig:cascade}). Furthermore, the spectral integration method applied on the cascade connection also produces erroneous results (not shown). This observation was made using both our implementation and implementation developed in~\cite{liejovJCP13}. 

	\vspace*{-4ex}
	\label{sec.2Dob}
\subsection{2D viscoelastic channel flow}\label{sec:apps2}

	\vspace*{-2ex}
As a second application, we consider 2D channel flow of an Oldroyd-B fluid described in \S~\ref{subsec:illcondDis}. In contrast to operator-induced ill-conditioning, discretization-induced errors can be alleviated by employing a well-conditioned discretization scheme, e.g., the ultraspherical and spectral integration schemes discussed in \S~\ref{sec:numMeth}. In conventional spectral methods (e.g., Chebyshev collocation method), discretization matrices become increasingly ill-conditioned with an increase in the number of basis functions. Viscoelastic channel flow requires a large number of basis functions for good resolution and provides an excellent benchmark for studying effects that arise from both discretization- and operator-induced ill-conditioning.

The frequency response operator $\MM T(\omega)$ for 2D channel flow of an Oldroyd-B fluid is described by~\eqref{eq:2Dvisco} and numerical implementation requires a large number of basis functions (about $4000$) for good resolution in a flow with moderate Weissenberg numbers ($\We \sim 50$). In strongly elastic flows (with $\We \sim 500$), an operator-induced ill-conditioning, similar to the one discussed \S~\ref{sec:apps1}, also arises. The discrete eigenvalues in a 2D flow scale as $1/\We$~\cite{renardy1986linear} and, at large $\We$, the cascade connection shown in~\fref{fig:cascade} is prone to ill-conditioning because of the inversions in $\MM T(\omega)\MM T^{\dagger}(\omega)$. At high elasticities, only the feedback connection in~Figure~\ref{fig:bd-fbk3} produces reliable results and all calculations in this section are based on it. 

	\vspace*{-3ex}
\subsubsection{Velocity output}

\begin{figure}
  \centering
  \subfigure[][Ultraspherical]{\includegraphics[scale = 0.21]{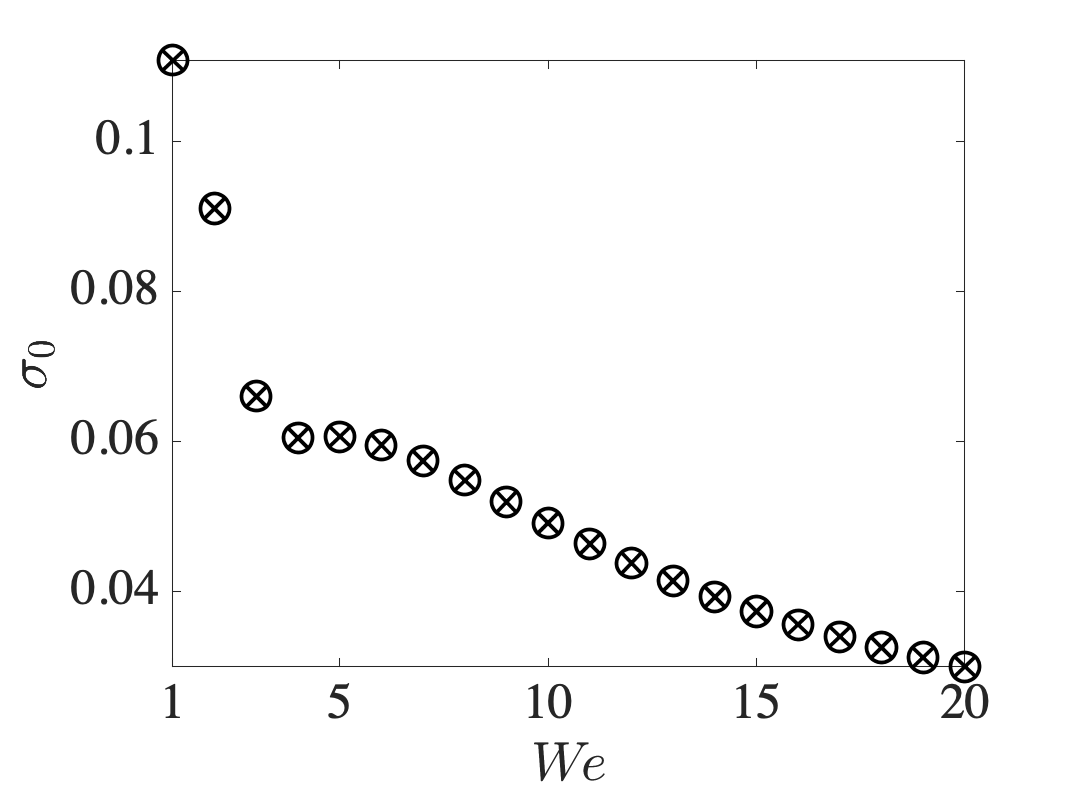}\label{fig:res2a}}
  \subfigure[][Spectral integration]{\includegraphics[scale = 0.21]{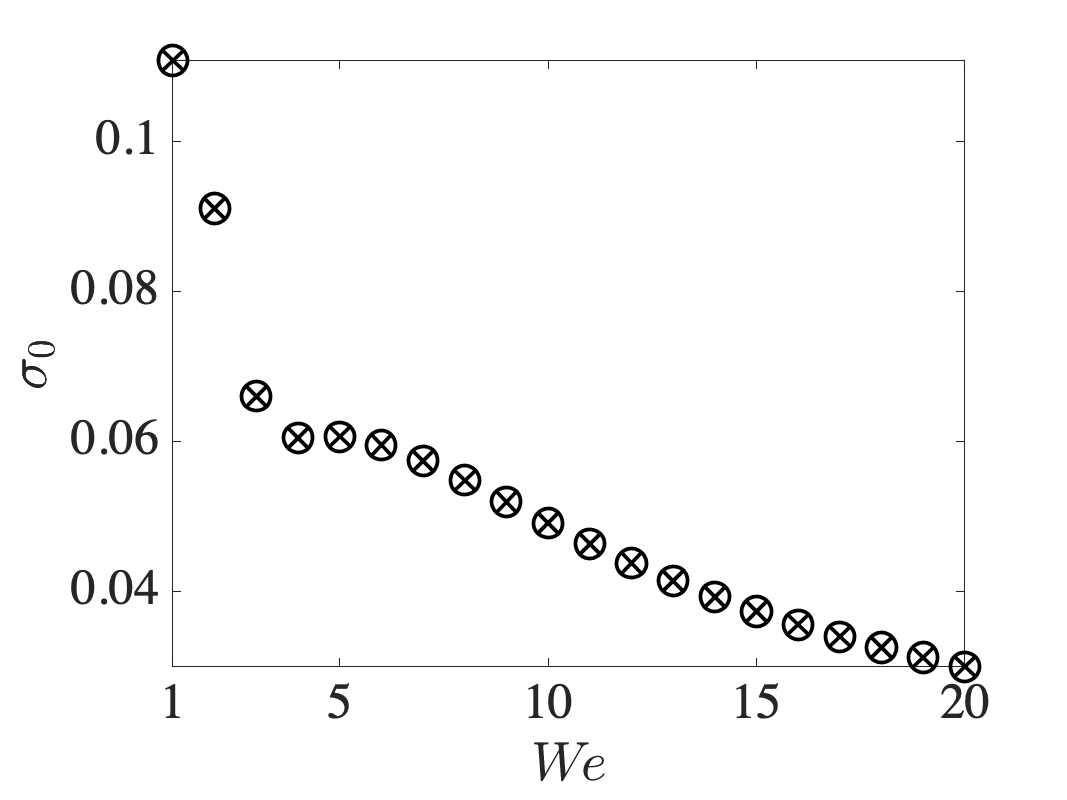}\label{fig:res2b}}
  \subfigure[][Spectral collocation]{\includegraphics[scale = 0.21]{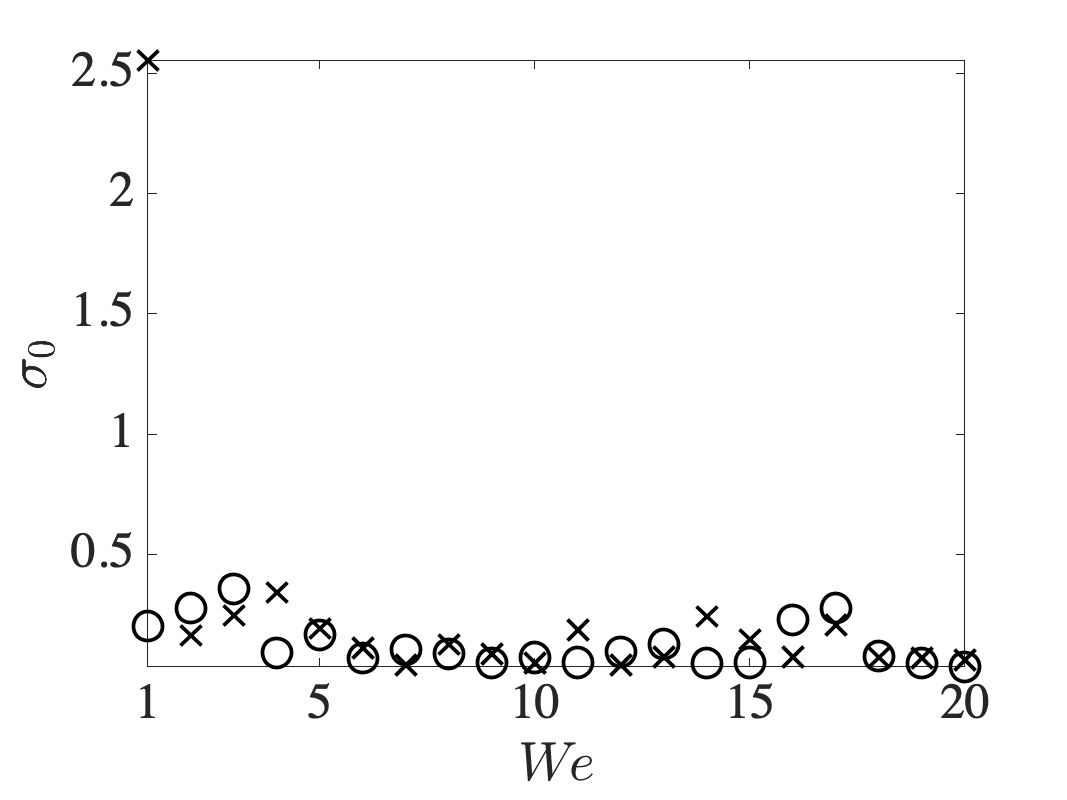}\label{fig:res2c}}
    \caption{\label{fig:res2} Principal singular values of the frequency response operator~\eqref{eq:2Dvisco} for inertialess 2D Couette flow of an Oldroyd-B fluid with $\beta = 0.5$, $k_x = 1$, and $\omega = 0$ as a function of fluid elasticity, $\We$, resulting from the use of (a) ultraspherical; (b) spectral integration; and (c) spectral collocation methods. The velocity fluctuations are selected as the output and symbols represent $N = 479$ ($\circ$) and $N = 511$ ($\times$).}
  \end{figure}

	\vspace*{-2ex}
For 2D Couette flow of an Oldroyd-B fluid, we employ spectral collocation, ultraspherical discretization, and spectral integration methods to compute singular values of the frequency response operator~\eqref{eq:2Dvisco} with the velocity as the output. In Figure~\ref{fig:res2}, we show the largest singular value as a function of $\We$ for $Re = 0$, $\beta = 0.5$, $k_x = 1$, and $\omega= 0$. Calculations are performed using $479$ (marked by circles) and $511$ (marked by crosses) basis functions. Figures \ref{fig:res2a} and \ref{fig:res2b} demonstrate that the ultraspherical and spectral integration methods produce grid-independent results. In contrast,~\fref{fig:res2c} illustrates that the spectral collocation method produces grid-dependent results.

In~\cite{liejovJCP13}, the performance of spectral integration and spectral collocation methods was compared using the same example. As in our study, it was observed that the collocation method produces unreliable, grid-dependent results, and that the spectral integration method yields reliable, grid-independent results. We find that the method based on ultraspherical discretization performs on par with the spectral integration method and that it produces grid-independent results for 2D Couette flow of an Oldroyd-B fluid with moderate $\We$.

\vspace*{-3ex}
\subsubsection{Stress output}

	\vspace*{-2ex}
When the stress fluctuations are selected as the output, we use the Chebfun's ultraspherical discretization in {\sf Matlab} for the frequency response analysis. The computations are verified using our spectral integration method (not reported here). Among other features, Chebfun offers the automatic collocation technique which increases the number of basis functions until the solution reaches machine precision~\cite{driscoll2014chebfun}. 

\begin{figure}
\centering
\subfigure[][]{\includegraphics[scale = 0.21]{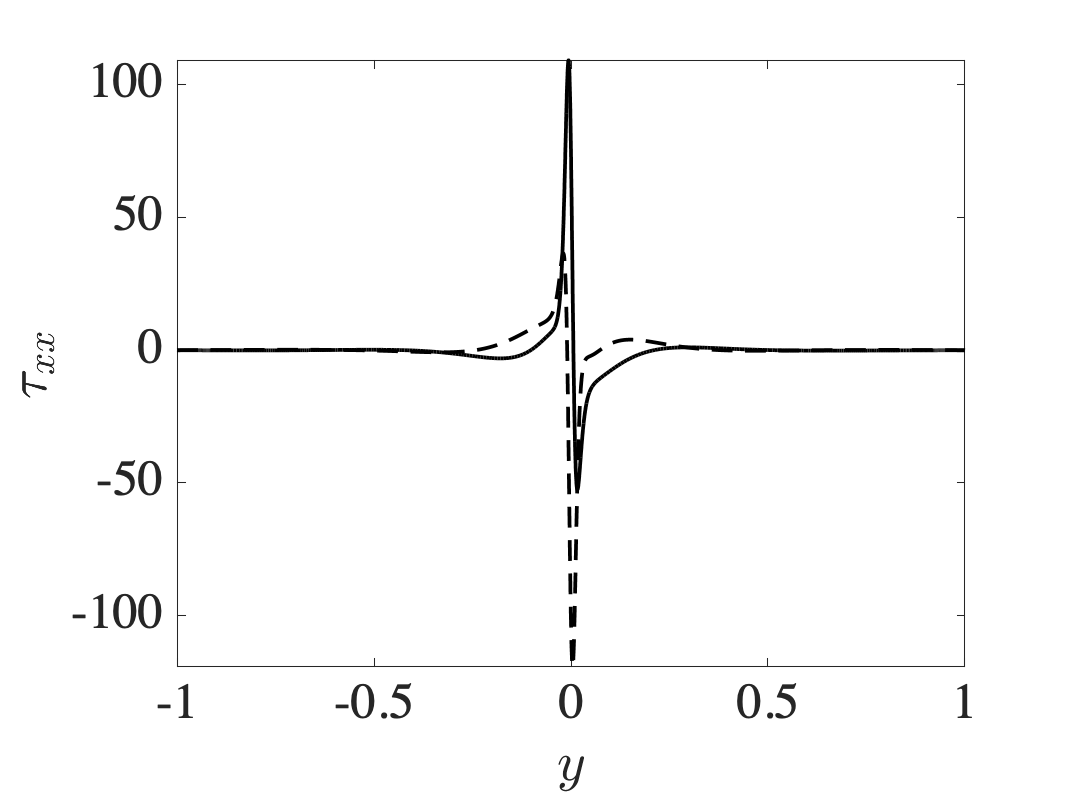}\label{fig:resEx3a}}
\subfigure[][]{\includegraphics[scale = 0.21]{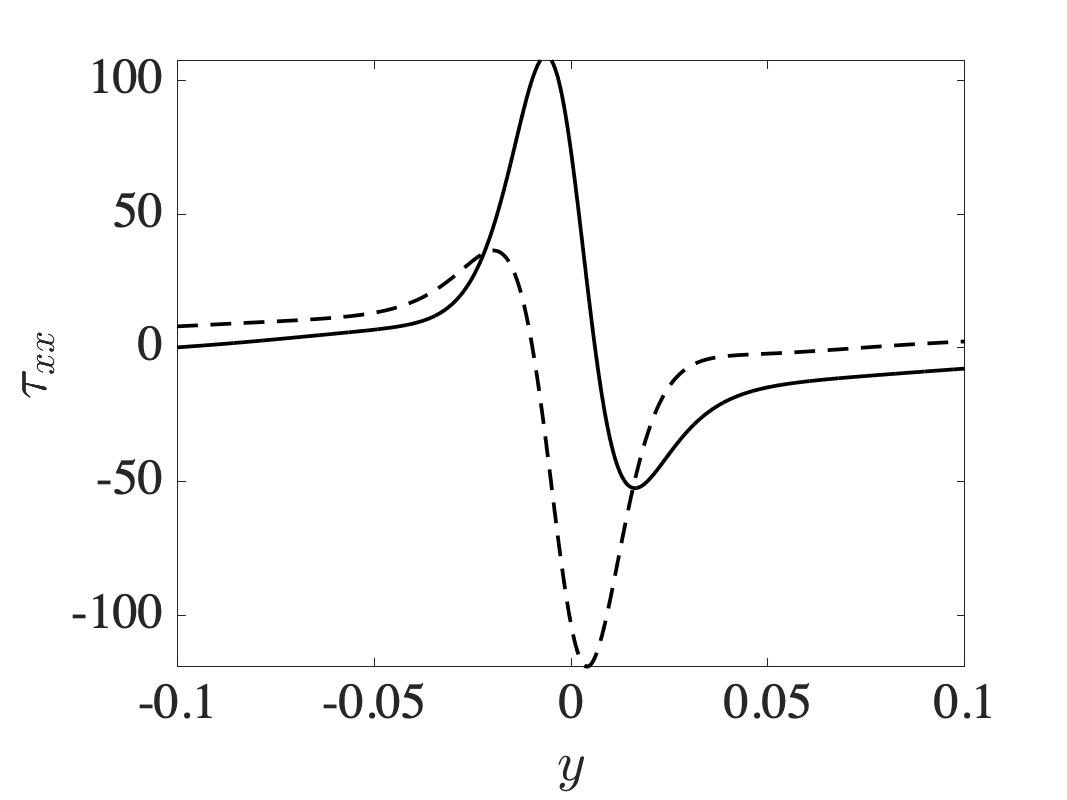}\label{fig:resEx3b}}
\caption{\label{fig:resEx3} The left singular function associated with the principal singular value $\sigma_{\max} = 14.936$ of inertialess 2D Couette flow of an Oldroyd-B fluid with $\We = 40$, $k_x = 1$, $\omega = 0$, and $\beta = 0.5$. The normal stress component, $\tau_{xx}$, (a) in the whole domain, $y\in[-1,1]$; and (b) near $y = 0$ is shown. The stress fluctuations are selected as the output and the lines correspond to $\re (\tau_{xx})$ (-), and $\im (\tau_{xx})$ (-\ -).}
	\vspace*{-0.5cm}
\end{figure}

The left singular functions associated with the largest singular value for the stress output reveal why these computations require a large number of basis functions. \fref{fig:resEx3} shows the principal left singular function of the normal stress component, $\tau_{xx}$, in inertialess 2D Couette flow with $\We = 40$, $\beta = 0.5$, $\omega = 0$, and $k_x = 1$. \fref{fig:resEx3a} illustrates $\tau_{xx}$ over the entire domain $y\in[-1,1]$, and~\fref{fig:resEx3b} shows $\tau_{xx}$ in the region where the highest values are achieved (near the center of the channel). In spite of large peak magnitudes, the left singular function is smooth and well-resolved.

\begin{figure}
\centering
\subfigure[][]{\includegraphics[scale = 0.21]{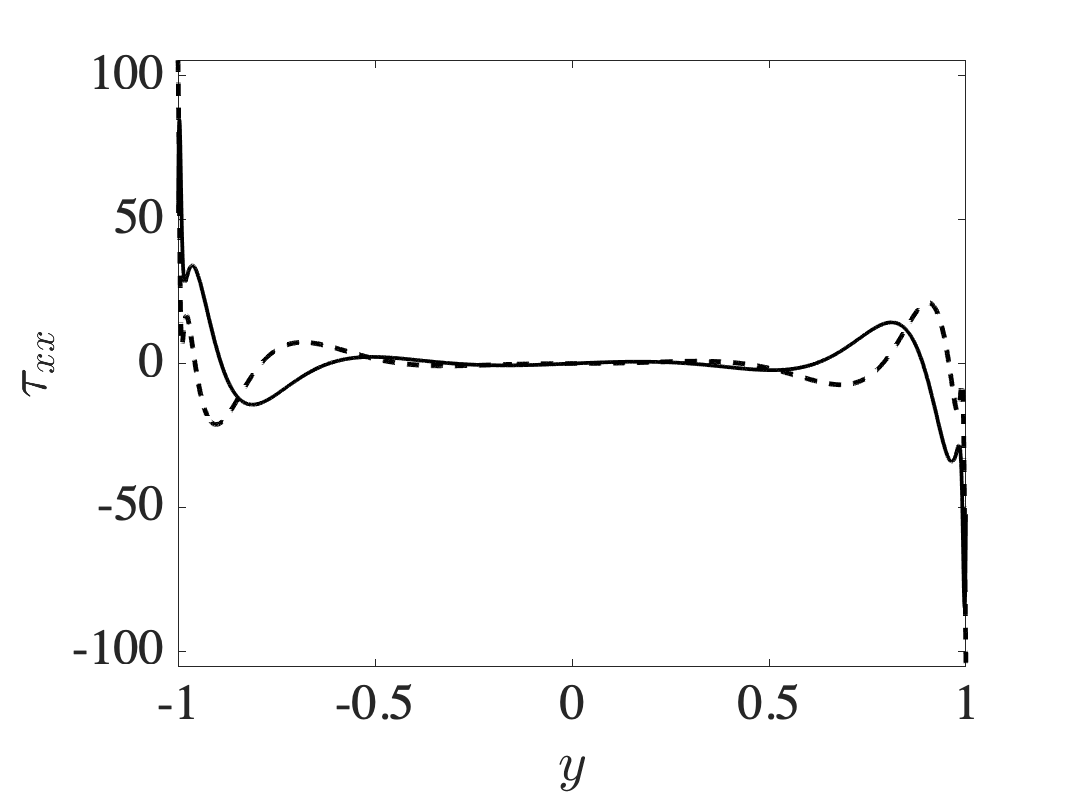}\label{fig:resEx5a}}
\subfigure[][]{\includegraphics[scale = 0.21]{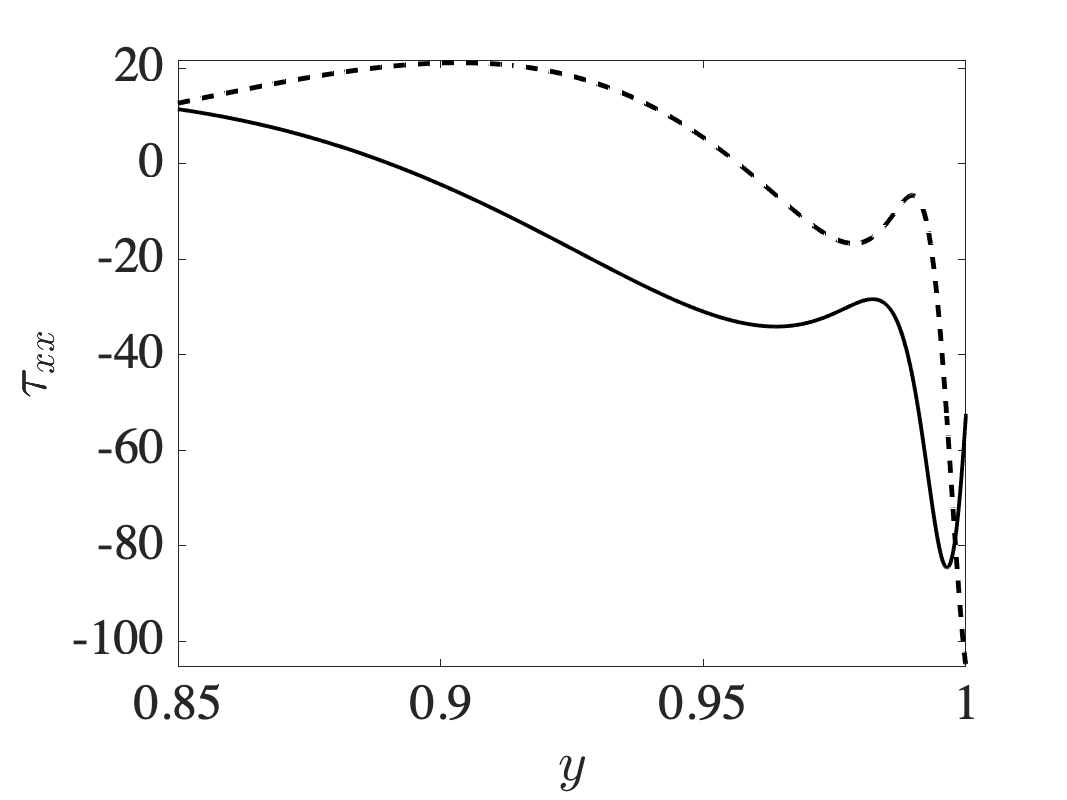}\label{fig:resEx5b}}
\caption{\label{fig:resEx5} The left singular function associated with the principal singular value $\sigma_{\max} = 6.184$ of  inertialess 2D Poiseuille flow of an Oldroyd-B fluid with $\We = 40$, $k_x = 1$, $\omega = 0$ and $\beta = 0.5$. The normal stress component, $\tau_{xx}$, (a) in the whole domain, $y\in[-1,1]$; and (b) near $y = 1$ is shown. The stress fluctuations are selected as the output and the lines correspond to $\re (\tau_{xx})$ (-), and $\im (\tau_{xx})$ (-\ -).}
\end{figure}

In contrast to the Couette flow computations, which require around $4000$ basis functions, the computations for Poiseuille flow were resolved to machine precision with around $1000$ basis functions. \fref{fig:resEx5} shows the principal left singular function for the stress output in Poiseuille flow that is obtained under the same conditions as~\fref{fig:resEx3} for Couette flow ($Re = 0$, $\We = 40$, $\beta = 0.5$, $k_x = 1$, and, $\omega = 0$). While in Couette flow the stress shows a steep variation near the channel center (see~\fref{fig:resEx3a}), in Poiseuille flow the steep variation occurs near the walls (see~\fref{fig:resEx5a}). Since interpolations based on Chebyshev polynomials utilize points that are more densely populated near the ends of the domain, sharp variations in Poiseuille flows can be resolved with a smaller number of basis functions than sharp variations in Couette flow. 

\begin{figure}
\centering
\subfigure[][]{\includegraphics[scale = 0.21]{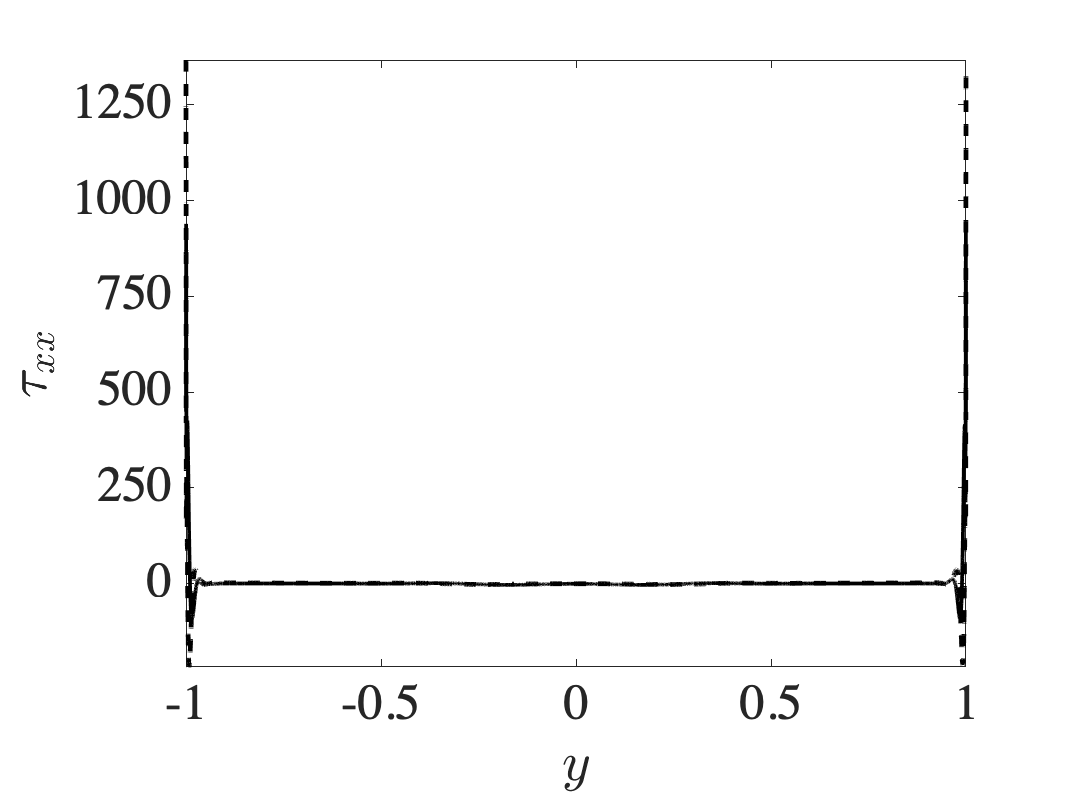}\label{fig:resEx6a}}
\subfigure[][]{\includegraphics[scale = 0.21]{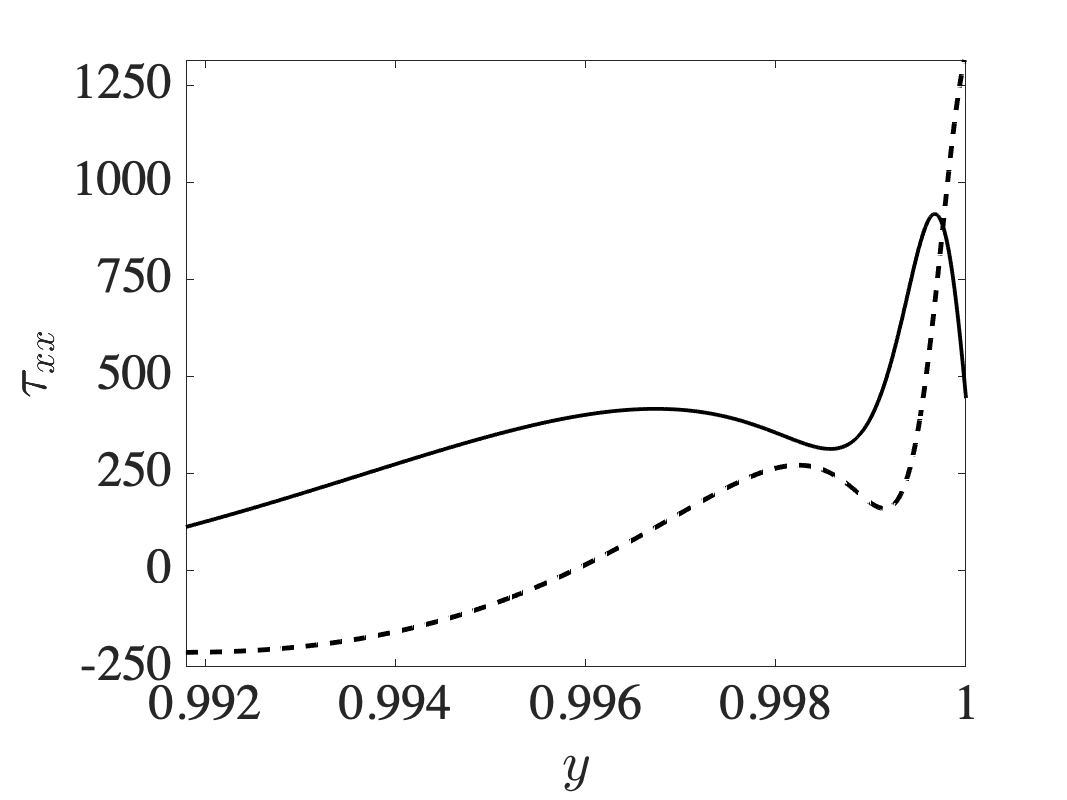}\label{fig:resEx6b}}\\
{\large $\sigma_0\;=\;5.98$}
\caption{\label{fig:resEx6} The left singular function associated with the principal singular value $\sigma_{\max} = 5.98$ of inertialess 2D Poiseuille flow of an Oldroyd-B fluid with $\We = 500$, $k_x = 1$, $\omega = 0$, and $\beta = 0.5$. The normal stress component, $\tau_{xx}$, (a) in the whole domain, $y\in[-1,1]$; and (b) near $y = 1$ is shown. The stress fluctuations are selected as the output and the lines correspond to $\re (\tau_{xx})$ (-), and $\im (\tau_{xx})$ (-\ -).}
	\vspace*{-0.5cm}
\end{figure}

Finally, we consider inertialess 2D Poiseuille flow with high elasticity ($\We = 500$), $\beta = 0.5$, $\omega = 0$, and $k_x = 1$. A well-resolved computation based on the feedback interconnection shown in Figure~\ref{fig:bd-fbk3} requires around $15000$ basis functions. We also used our implementation of the spectral integration method (described in~\sref{sec:numMeth}) as well as the spectral integration code developed in~\cite{liejovJCP13} to verify that the approach based on a cascade connection shown in \fref{fig:cascade} fails to produce reliable results. The principal left singular function corresponding to $\tau_{xx}$ is shown in \fref{fig:resEx6}. As expected, steep variations near $y = \pm 1$ are observed with the peak value of around $1000$. \fref{fig:resEx6b} shows a close-up of \fref{fig:resEx6a} near $y = 1$ and demonstrates that the most amplified output direction is well-resolved even though the variation in $\tau_{xx}$ is spanning three orders in magnitude within the region of width $10^{-3}$ in $y\in [-1, 1]$. \tcp{The computed principal singular value, $\sigma_0 = 5.98$, has an imaginary part of $\mathcal{O}(10^{-7})$ and the corresponding singular function is resolved to machine accuracy as verified by the Chebfun's automatic collocation procedure~\cite{OlvTowSIAM2013,driscoll2010automatic,driscoll2014chebfun}.}

	\vspace*{-4ex}
\subsection{Frequency response analysis of systems in the descriptor form}\label{sec:apps3}

  \vspace*{-2ex}
For 2D viscoelastic fluids in the evolution form, \S~\ref{sec.2Dob} demonstrates that both the ultraspherical and the spectral integration methods produce reliable results. We next utilize the formulation based on the feedback interconnection shown in Figure~\ref{fig:bd-fbk3} in conjunction with the spectral integration method for frequency response analysis of systems in the descriptor form. We examine the linearized NS equations presented in~\sref{sec:MotEx3} and the 3D flow of an Oldroyd-B fluid with the stress fluctuations eliminated~(see \sref{sec:desStressElim}). As discussed in \S~\ref{sec:MotEx3}, for incompressible flows in the descriptor form, conventional spectral methods require a staggered grid which may be difficult to implement in generic solvers like Chebfun~\cite{driscoll2014chebfun}. Our spectral integration method overcomes this challenge by reinforcing the algebraic constraint~\eqref{eq:lnsb} at the walls; see~\S~\ref{sec:bcs}.

	\vspace*{-4ex}
\subsubsection{Channel flow of a Newtonian fluid}\label{sec:app3a}

\begin{figure}
  \centering
  \subfigure[][Modal]{\includegraphics[scale = 0.21]{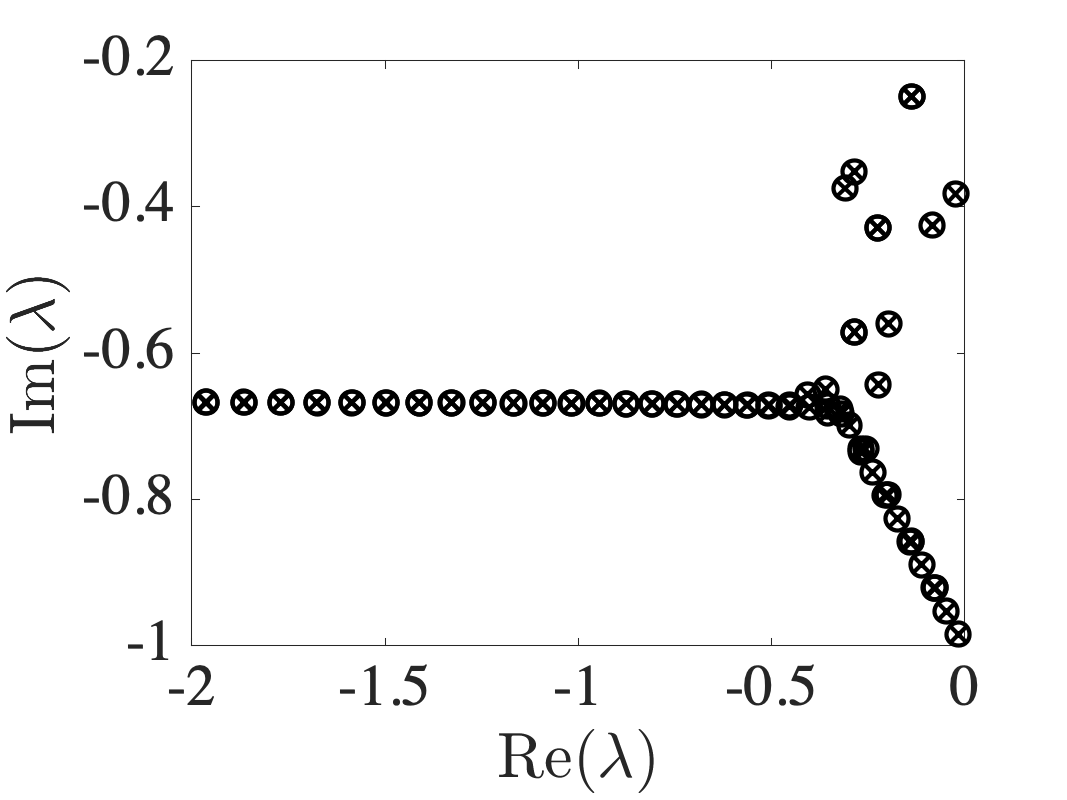}\label{fig:res3a}}
  \subfigure[][Nonmodal]{\includegraphics[scale = 0.21]{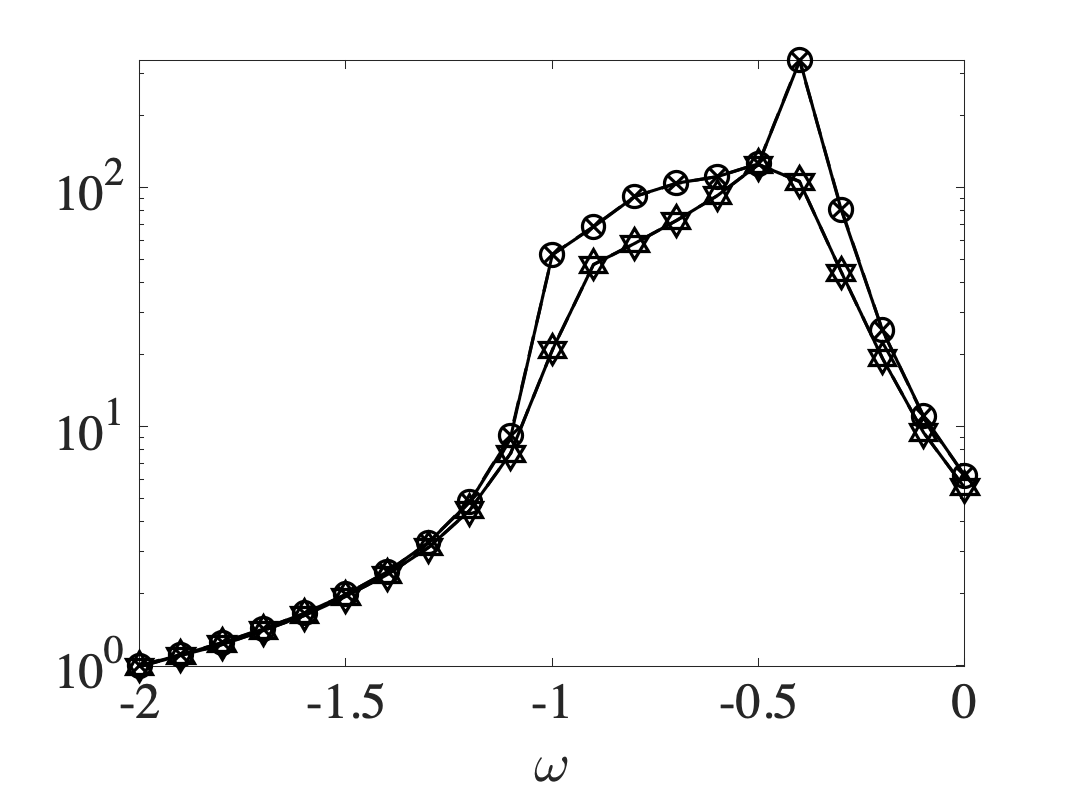}\label{fig:res3b}}
    \caption{\label{fig:res3} The linearized NS equations in Poiseuille flow with $Re = 2000$ and $k_x = k_z = 1$. The spectral integration method with $N = 255$ basis functions is used. (a) Spectrum resulting from the use of the evolution form model ($\times$) and the descriptor formulation ($\circ$); and (b) two largest singular values of the frequency response operator (evolution form ($\times$) and descriptor formulation ($\circ$) results for $\sigma_{\max}$; evolution form ($\bigtriangleup$) and descriptor formulation ($\bigtriangledown$) results for the second largest singular value). \tcg{The singular values in (b) agree with the values reported in~\cite[Figure 4.10]{schmid2012stability}.}}  
    \end{figure}

	\vspace*{-2ex}
We first examine the linearized NS equations in Poiseuille flow; see~\eqref{eq:lns} and \fref{fig:flow_geo} for geometry. Modal analysis considers temporal growth or decay of infinitesimal fluctuations around the parabolic velocity profile $U (y) = 1-y^2$. For $Re = 2000$, the linearized NS equations are stable~\cite{schmid2012stability} and \fref{fig:res3a} shows the spectrum of the flow with $k_x = k_z = 1$. The results are obtained using the spectral integration method with $255$ basis functions. \fref{fig:res3a} shows that all eigenvalues are in the left-half of the complex plane and demonstrates the absence of spurious modes. We note that the computations based on the evolution form model (crosses) and the descriptor formulation (circles) agree with each other and with the results reported in the literature~\cite{schmid2012stability}. 

\fref{fig:res3b} shows the dependence on the temporal frequency of the two largest singular values of the frequency response operator. For the principal singular value, the evolution form model results are marked by crosses and the descriptor formulation results are marked by circles. For the second largest singular value, the evolution form model results are marked by triangles and the descriptor formulation results are marked by inverted triangles. We observe excellent agreement in both cases.

  	\vspace*{-4ex}
\subsubsection{Channel flow of an Oldroyd-B fluid}\label{sec:app3b}
\begin{figure}
  \centering
  \subfigure[]{\includegraphics[scale = 0.21]{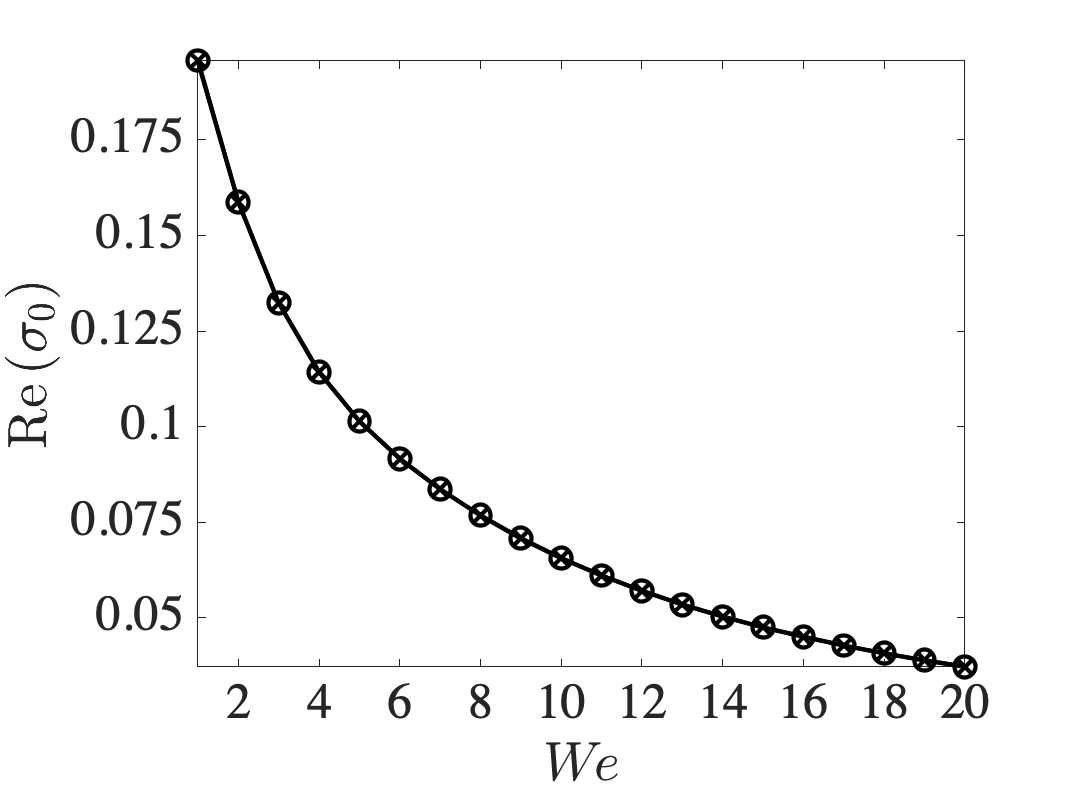}\label{fig:visDesa}}
  \subfigure[]{\includegraphics[scale = 0.21]{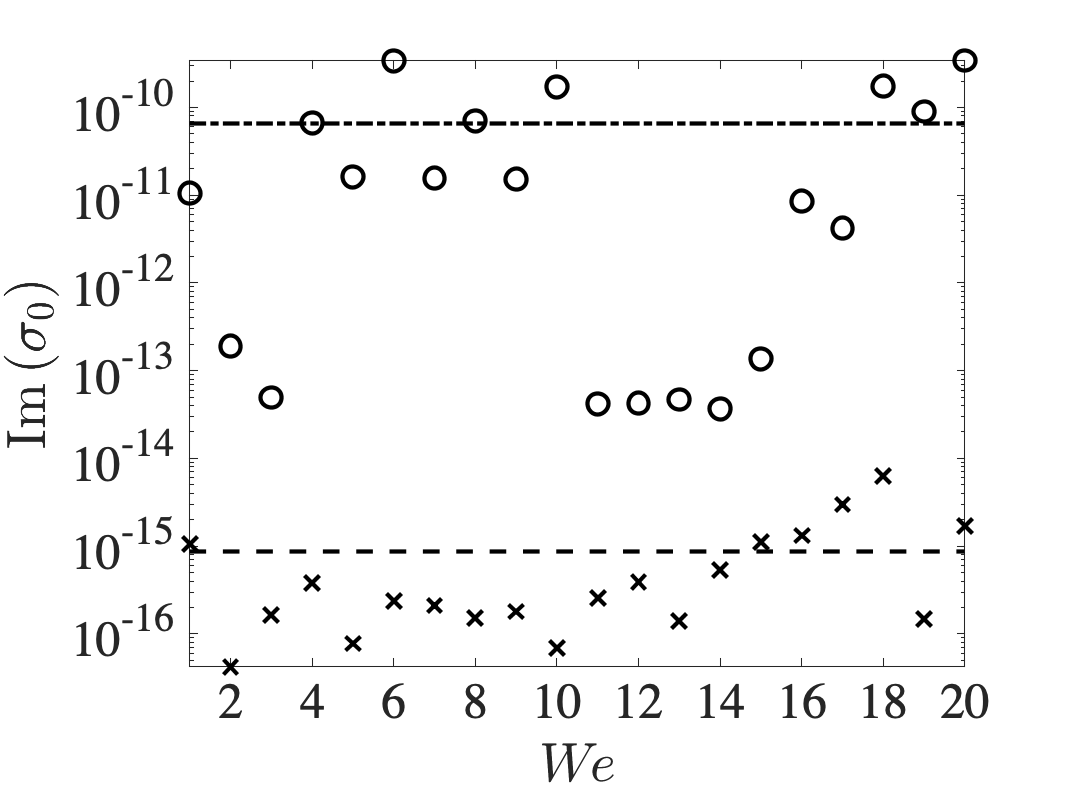}\label{fig:visDesb}}
    \caption{\label{fig:visDes} (a) Real; and (b) imaginary parts of the principal singular value in inertialess Couette flow of an Oldroyd-B fluid with $\beta = 0.5$, $k_x = k_z = 1$, and $\omega = 0$. The velocity fluctuations are selected as the output and the results are obtained using the descriptor formulation $(\times)$ that eliminates stresses (see~\sref{sec:desStressElim}) with $N = 383$ basis functions and the evolution form model $(\circ)$ (see~\sref{sec:evolForm}) with $N = 1000$ basis functions.}
    	\vspace*{-0.5cm}
  \end{figure}
      
     \vspace*{-2ex}
\fref{fig:visDesa} demonstrates the agreement between the singular values obtained using the descriptor formulation ($\times$) with $N = 383$ basis functions and the evolution formulation ($\circ$) with $N = 1000$ basis functions. For inertialess Couette flow of an Oldroyd-B fluid with $\beta = 0.5$, $k_x = k_z = 1$, and $\omega = 0$, the velocity fluctuations are selected as the output and the influence of fluid elasticity $\We$ on the principal singular value is shown. Although the imaginary part of a computed singular value should be equal to zero, its value depends on the accuracy of the numerical method, and a smaller imaginary part signals higher accuracy. \fref{fig:visDesb} displays the imaginary part of the principal singular value; the average imaginary part resulting from the descriptor formulation (with $N = 383$) and from the evolution formulation (with $N = 1000$) are, respectively, determined by $\sim 10^{-15}$ (dashed line) and $\sim 10^{-10}$ (dashed dotted line). \tcp{Imaginary parts in~\fref{fig:visDesb} are obtained by increasing the number of basis functions until no further decrease is observed and they are of $\mathcal{O} (10^{-10})$ and $\mathcal{O} (10^{-15})$ for the evolution and descriptor form models, respectively.}

\begin{figure}
  \centering
  \subfigure[]{\includegraphics[scale = 0.21]{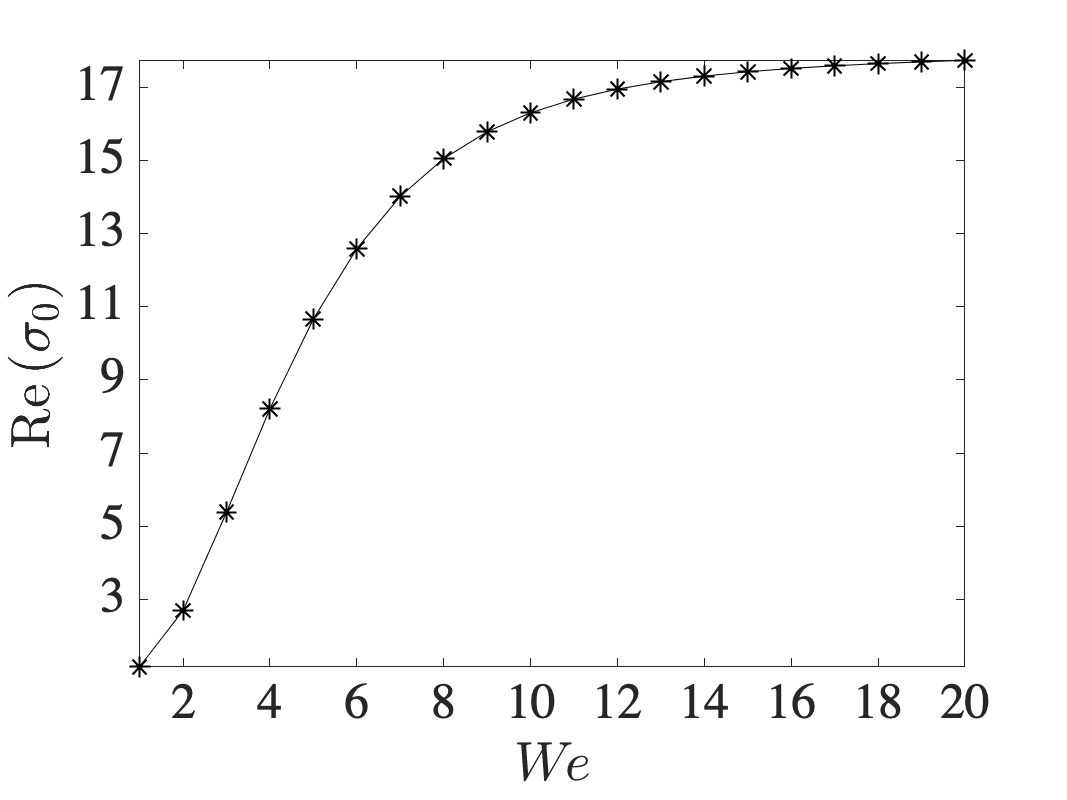}\label{fig:visDesSa}}
  \subfigure[]{\includegraphics[scale = 0.21]{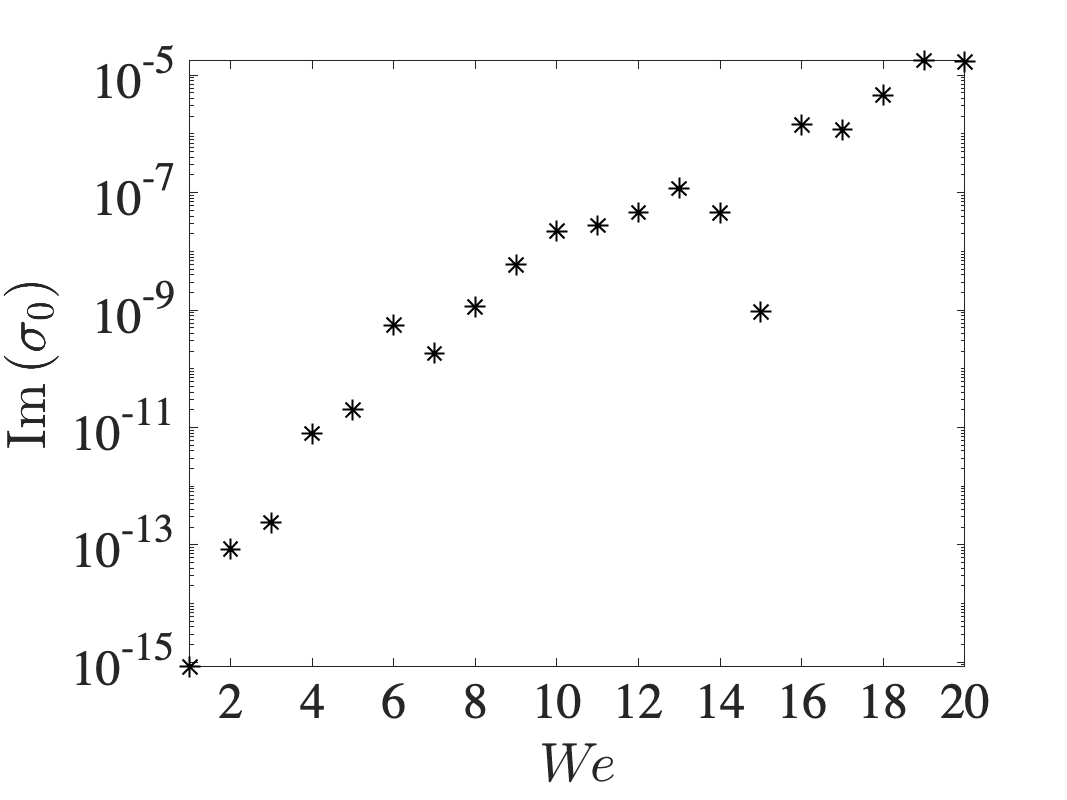}\label{fig:visDesSb}}
    \caption{\label{fig:visDesS} (a) Real; and (b) imaginary parts of the principal singular value in inertialess Couette flow of an Oldroyd-B fluid with $\beta = 0.5$, $k_x = k_z = 1$, and $\omega = 0$. The normal stress component, $\tau_{xx}$, is selected as the output and the results are obtained using the descriptor formulations that eliminates stresses (see~\sref{sec:desStressElim}) with $N = 863$ basis functions.}
    \vspace*{-0.375cm}
  \end{figure}

\begin{figure}
  \centering
  \subfigure[][]{\includegraphics[scale = 0.21]{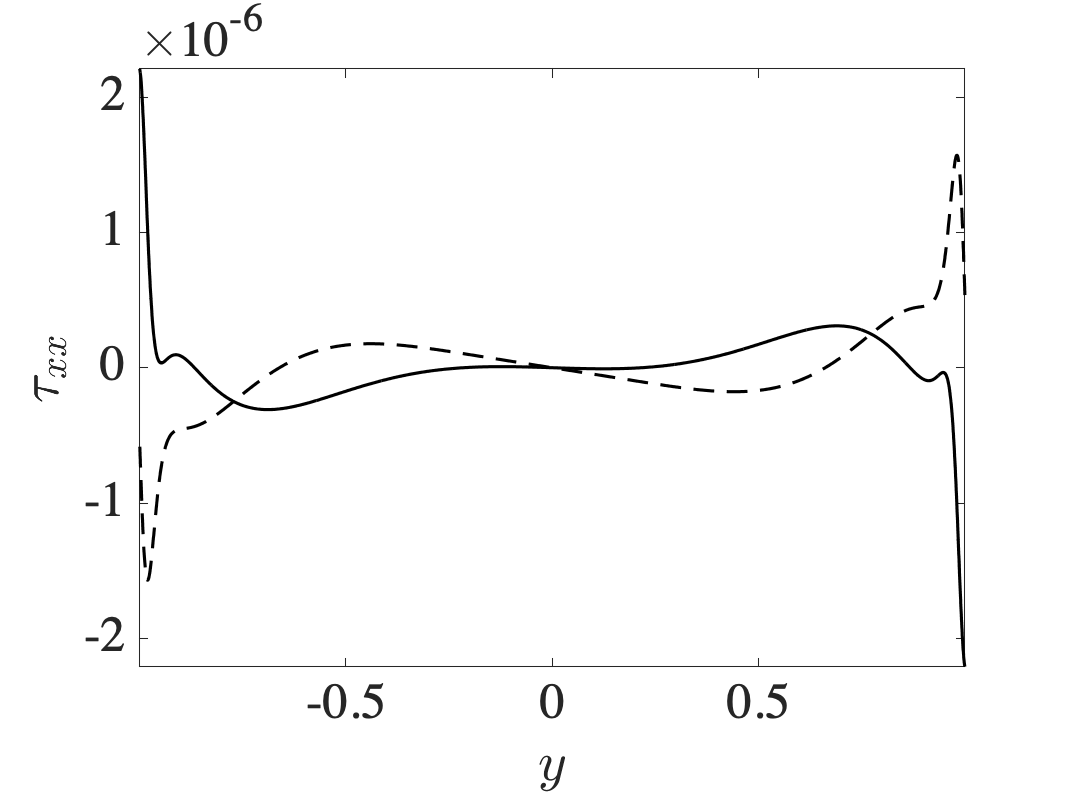}\label{fig:visDesFa}}
  \subfigure[][]{\includegraphics[scale = 0.21]{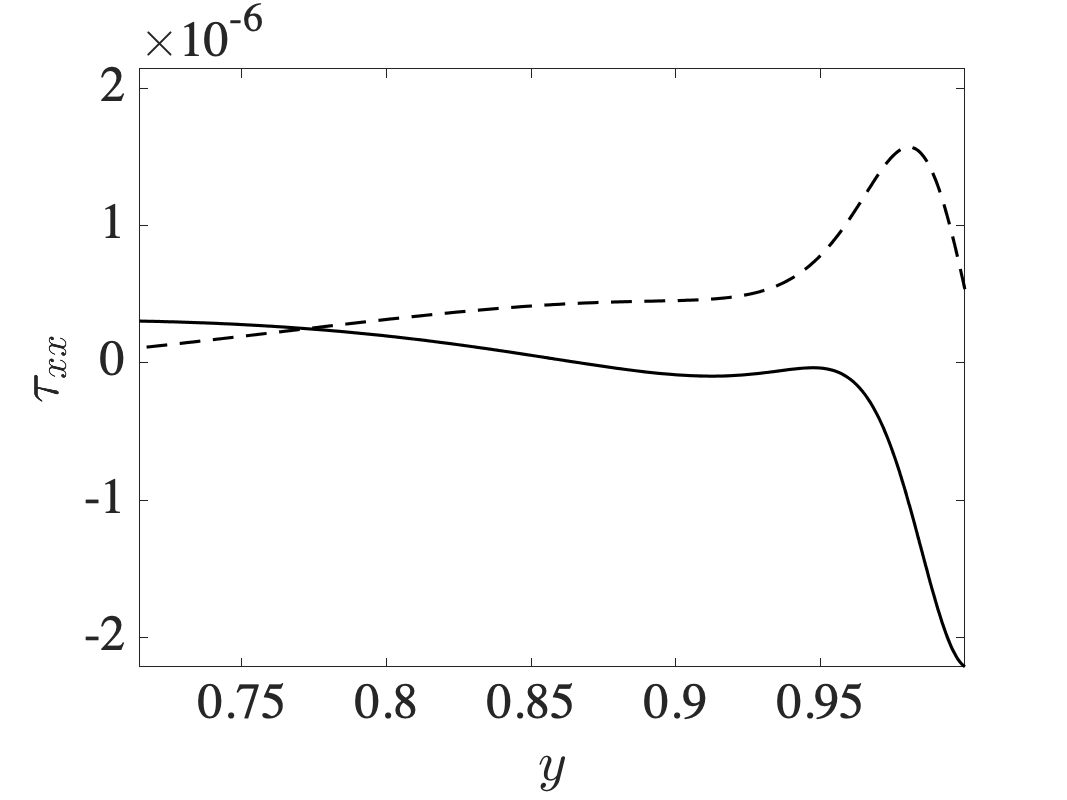}\label{fig:visDesFb}}\\
    \caption{\label{fig:visDesF} The left singular function corresponding to the principal singular value $\sigma_{\max} = 7.434$ in inertialess Poiseuille flow of an Oldroyd-B fluid with $\We = 10$, $\beta = 0.5$, $k_x = k_z = 1$, and $\omega = 0$. The normal stress component, $\tau_{xx}$, is selected as the output and the results are obtained using the descriptor formulations that eliminates stresses with $N = 863$ basis functions. The lines correspond to $\re (\tau_{xx})$ (-) and $\im (\tau_{xx})$ (-\ -) and the results (a) in the entire domain, $y\in[-1,1]$; and (b) near $y = 1$, are shown.}
    \vspace*{-0.5cm}
  \end{figure}

For inertialess Couette flow of an Oldroyd-B fluid with $\beta = 0.5$, $k_x = k_z = 1$, and $\omega = 0$, \fref{fig:visDesSa} shows the $\We$-dependence of the principal singular value of the frequency response operator. The normal stress component, $\tau_{xx}$, is selected as the output and the computations are obtained using the descriptor formulation with $N = 863$ basis functions. The principal singular value increases with fluid elasticity but it appears to saturate for large values of $\We$. For fixed $N$, \fref{fig:visDesSb} demonstrates that the imaginary part of the principal singular value becomes larger with an increase in $\We$. We further observe that for $\We > 20$, the accuracy of the computed singular values does not improve with a further increase in $N$ beyond certain value ($\approx 863$) and that the frequency response analysis of plane Poiseuille flow with stress as the output shows similar trends (not shown).

For inertialess Poiseuille flow with $\We = 10$, $\beta = 0.5$, $k_x = k_z = 1$, and $\omega = 0$, \fref{fig:visDesF} shows the principal left singular function of the frequency response operator with $\tau_{xx}$ as the output. The descriptor formulation is used in our computations and, as in 2D flow, we observe sharp stress gradients near the walls. 

	\vspace*{-3ex}
\section{Concluding remarks}\label{sec:summary}

	\vspace*{-2ex}
In this paper, we explore the merits and the effectiveness of well-conditioned ultraspherical and spectral integration methods for nonmodal analysis of channel flows of Newtonian and viscoelastic fluids. We develop a framework for resolvent analysis that is based on a feedback interconnection of the frequency response operator with its adjoint and demonstrate its advantages over the standard formulation that utilizes a cascade connection. For ill-conditioned problems, we show that a combination of the formulation based on this feedback interconnection with well-conditioned ultraspherical and spectral integration methods can be used to overcome limitations of standard spectral collocation techniques. In particular, we demonstrate that our approach provides robust results in channels flows of Oldroyd-B fluids with high elasticity and show that the spectral integration method does not require a staggered grid for modal or nonmodal analysis of channel flows of incompressible fluids in the descriptor form. This facilitates analysis of relevant flow physics in strongly elastic regimes and enables computations using the formulation with primitive variables. For a given number of basis functions, we show that the computations resulting from the descriptor formulation are more accurate than the computations based on the evolution formulation. Even though we focus on nonmodal analysis of channel flows of Newtonian and viscoelastic fluids, the developed framework is general enough to find use for a variety of problems in fluid mechanics and beyond. 

\vspace*{-3ex}
\section*{Software}
	\vspace*{-2ex}

Additional information about the spectral integration method along with source codes for A Matlab Spectral Integration Suite ({\sf SISMatlab}) that we developed can be found at: 
	\begin{center}
	\href{https://viterbi-web.usc.edu/~mihailo/software/sismatlab/}{https://viterbi-web.usc.edu/$\sim$mihailo/software/sismatlab/}
	\end{center}

	\vspace*{-4ex}
\section*{Acknowledgments}

	\vspace*{-2ex}
This work is supported in part by the National Science Foundation under Award CBET-1510654 and by the Air Force Office of Scientific Research under Award FA9550-18-1-0422. The Minnesota Supercomputing Institute (MSI) at the University of Minnesota is acknowledged for providing computing resources.

\vspace*{-4ex}
\singlespacing

\newpage

\appendix
\section{Operators governing 2D channel flow of an Oldroyd-B fluid}\label{appA}

	\vspace*{-2ex}
The variable coefficients for the operator in \eqref{eq:2Dvisco} are given by (we have suppressed the dependence on $\omega$ for the sake of brevity, e.g., $a_0(y,\omega)$ is denoted as $a_0(y)$)
\begin{align*}
  a_4(y)  \;=&\; \frac{\beta -1}{c(y)}-\beta,\\
  a_3(y)  \;=&\; -\frac{2 (\beta -1) \left(c'(y)-\mri \,k_x\, \We c(y) U'(y)\right)}{c(y)^2},\\
  \begin{split}
      a_2(y)  \;=&\; \frac{\,(1-\beta)\, c''(y)}{c(y)^2}+\frac{2 \mri \,(1-\beta)\, \,k_x \,\We c'(y) U'(y)}{c(y)^2}\\
      &-\frac{2 \,(1-\beta)\, \,k_x^2
      \,\We^2 U'(y)^2}{c(y)^2}+\frac{2 \mri \,(1-\beta)\, \,k_x \,\We U''(y)}{c(y)^2}\\
      &-\frac{4 \mri \,(1-\beta)\, \,k_x \,\We c'(y)
   U'(y)}{c(y)^3}-\frac{2 \,(1-\beta)\, c'(y)^2}{c(y)^3}-\frac{4 \,(1-\beta)\, \,k_x^2 \,\We^2 U'(y)^2}{c(y)^3}\\
   &+\frac{2 \,(1-\beta)\, \,k_x^2 \,\We^2 U'(y)^2}{c(y)}+\frac{2 \,(1-\beta)\, \,k_x^2}{c(y)}-\frac{3 \mri
   \,(1-\beta)\, \,k_x \,\We U''(y)}{c(y)}+2 \beta  \,k_x^2,
   \end{split}\\
   \begin{split}
        a_1(y)  \;=&\;\frac{12
   \,(1-\beta)\, \,k_x^2 \,\We^2 c'(y) U'(y)^2}{c(y)^4}+\frac{12 \mri \,(1-\beta)\, \,k_x \,\We c'(y)^2
   U'(y)}{c(y)^4}\\
   &-\frac{4 \mri \,(1-\beta)\, \,k_x \,\We c''(y) U'(y)}{c(y)^3}+\frac{8 \,(1-\beta)\, \,k_x^2 \,\We^2 c'(y) U'(y)^2}{c(y)^3}\\
   &-\frac{8 \mri \,(1-\beta)\, \,k_x \,\We c'(y) U''(y)}{c(y)^3}-\frac{4 \mri \,(1-\beta)\, \,k_x^3 \,\We^3
   U'(y)^3}{c(y)^3}\\
   &-\frac{8 \,(1-\beta)\, \,k_x^2 \,\We^2 U'(y) U''(y)}{c(y)^3}\\
   &-\frac{2 \,(1-\beta)\, \,k_x^2 c'(y)}{c(y)^2}+\frac{2 \mri \,(1-\beta)\, \,k_x \,\We
   c'(y) U''(y)}{c(y)^2}\\
   &-\frac{4 \mri \,(1-\beta)\, \,k_x^3 \,\We^3 U'(y)^3}{c(y)^2}-\frac{6
   \,(1-\beta)\, \,k_x^2 \,\We^2 U'(y) U''(y)}{c(y)^2}\\
   &+\frac{2 \mri \,(1-\beta)\, \,k_x^3 \,\We U'(y)}{c(y)}+\frac{4 \,(1-\beta)\, \,k_x^2 \,\We^2 U'(y) U''(y)}{c(y)},
   \end{split}
\end{align*}
where $c(y) = i\,\omega + 1 + \mri\,k_x\,\We\, U(y)$,
\begin{align*}
  c_{11}(y)  \;=&\; \frac{2 \,\We U'(y)}{c(y)}+\frac{2 \,\We U'(y)}{c(y)^2},\\
  c_{12}(y)  \;=&\;  \frac{4 \mri \,k_x \,\We^2 U'(y)^2}{c(y)}-\frac{4\mri \,k_x \,\We^2 U'(y)^2}{c(y)^3}+\frac{2 \mri \,k_x}{c(y)},\\
      c_{13}(y)  \;=&\; +\frac{4 \,k_x^2 \,\We^3 U'(y)^3}{c(y)^3} \\
      &\;+ \frac{4 \,k_x^2 \,\We^3 U'(y)^3}{c(y)^2}+\frac{2 \,k_x^2 \,\We U'(y)}{c(y)^2}+\frac{2 i \,k_x \,\We^2 U'(y) U''(y)}{c(y)^2}\\
      &\;+\frac{4 \mri \,k_x \,\We^2
   U'(y) U''(y)}{c(y)},\\
   c_{21}(y)  \;=&\; \frac{1}{c(y)},
\end{align*}\newpage
\noindent \begin{align*}
  c_{22}(y)  \;=&\; -\frac{2 \mri \,k_x \,\We U'(y)}{c(y)^2},\\
   c_{23}(y)  \;=&\; \frac{2 \,k_x^2 \,\We^2 U'(y)^2}{c(y)}+\frac{2 \,k_x^2 \,\We^2 U'(y)^2}{c(y)^2}+\frac{\,k_x^2}{c(y)}+\frac{\mri \,k_x \,\We U''(y)}{c(y)},\\
   c_{31}(y)  \;=&\; -\frac{2 \mri \,k_x}{c(y)},\\
   c_{32}(y)  \;=&\; \frac{2 \,k_x^2 \,\We U'(y)}{c(y)}.
\end{align*}

	\vspace*{-4ex}
\section{Operators governing 3D channel flow of an Oldroyd-B fluid}\label{sec:appc}

	\vspace*{-2ex}
The nonzero components in the operator $\mathcal V $ in \eqref{eq:tauTov} are derived from the following relations that come from \eqref{eq:2.1c}. Note that $c(y) = i\,\omega + 1/\We + \mri\,k_x\, U(y)$ in this section.
\begin{subequations}\label{eq:tauRels}
\begin{align}
  \tau_{zz}(y) \;&=\; \frac{2\,\mri\,k_z}{\We \,c(y)}\,w(y)\\
  \tau_{yz}(y) \;&=\; \frac{1}{c(y)}\left(\frac{\mri\,k_z}{\We}\, v(y)+ \mri\,k_x\,T_{xy}(y)\, w(y) + \frac{w'(y)}{\We}\right)\\
  \tau_{yy}(y) \;&=\; \frac{2\,\mri\,k_x\,T_{xy}(y)}{c(y)}\,v(y) + \frac{2}{\We\,c(y)}\,v'(y)\\
  \tau_{xz}(y) \;&=\; \frac{1}{c(y)}\left(U'(y)\tau_{yz}(y) + \mri\,k_x\,T_{xx}(y)\,w(y) + \frac{\mri\,k_z}{\We} u(y) + \frac{\mri\,k_x}{\We}w(y) + T_{xy}(y) w'(y)\right)\\
  \begin{split}
    \tau_{xy}(y)\;&=\; \frac{1}{c(y)}\left(\mri\,k_x\,T_{xy}(y) u(y)+ \mri\,k_x\,T_{xx}(y) v(y) - T_{xy}'(y) v(y)\right) + \\ &+ \frac{1}{c(y)}\left(\frac{\mri\,k_x}{\We}v(y) + \frac{1}{\We}u'(y) + U'(y)\tau_{yy} + T_{xy}(y)\, v'(y)\right)
    \end{split}\\
  \tau_{xx}(y) \;&=\; \frac{1}{c(y)}\left(\frac{2\,\mri\,k_x}{\We}u(y) + 2\,k_x\,T_{xx}(y)\,u(y) - T_{xx}'(y)\,v(y) + 2 T_{xy}(y) u'(y) + 2\,U'(y)\tau_{xy}(y)\right)
\end{align}
\end{subequations}

	\vspace*{-4ex}
  \subsection{Evolution form}\label{sec:appCEvol}
  
  	\vspace*{-2ex}
  The state variables for this system are the wall-normal velocity and vorticity, $\bphi = [\,v ~ \eta\,]^T$ in~\eqref{eq:regularOp}. The boundary conditions are given by
  \begin{align}\label{eq:vortBc}
    v(\pm 1) \;=\; [\DD v (\cdot)](\pm1)\;=\; \eta (\pm 1) \;=\; 0.
  \end{align}
  The operator-valued matrices $\MM A$, $\MM B$, and $\MM C$ in \eqref{eq:mot1} are presented in this section. $\MM A$ is of size $2\times 2$ with 
  \begin{subequations}
    \begin{align*}
      \MM A_{11} \;&=\; \sum_{n \, = \, 0}^4 a_{n,11}(y,\omega) {\mathrm D}^n,
      \\
      \MM A_{12} \;&=\; 0,
      \\
      \MM A_{21} \;&=\; \sum_{n \, = \, 0}^2 a_{n,21} (y,\omega)\HH {\mathrm D}^n,
      \\
      \MM A_{22} \;&=\; \sum_{n \, = \, 0}^2 a_{n,22} (y,\omega) {\mathrm D}^n,
    \end{align*}
    where the nonzero elements of $\MM A$ are given by\newpage
    \noindent 
    \begin{align*}
      a_{4,11} \;=&\; -\frac{(1-\beta)\, }{\We\, c(y)}-\beta,\\ 
 a_{3,11} \;=&\; \frac{2 (1-\beta)\,  c'(y)}{\We\, c(y)^2}-\frac{2 \mri (1-\beta)\,  k_x\, T_{xy}(y)}{c(y)},\\ 
a_{2,11} \;=&\; \frac{(1-\beta)\,  c''(y)}{\We\, c(y)^2}+\frac{2 \mri (1-\beta)\,  k_x\, T_{xy}(y) c'(y)}{c(y)^2}-\frac{4 \mri (1-\beta)\,  k_x\, c'(y) U'(y)}{\We\, c(y)^3}-\frac{2 (1-\beta)\,  c'(y)^2}{\We\, c(y)^3}\\
&\;+\frac{2 (1-\beta)\,  k^2\,}{\We\, c(y)}+\frac{(1-\beta)\,  k_x^2\, T_{xx}(y)}{c(y)}-\frac{2 (1-\beta)\,  k_x^2\, T_{xy}(y) U'(y)}{c(y)^2}-\frac{4 (1-\beta)\,  k_x^2\, U'(y)^2}{\We\, c(y)^3}\\
&\;-\frac{3\mri (1-\beta)\,  k_x\, T_{xy}'(y)}{c(y)}+\frac{2 \mri (1-\beta)\,  k_x\, U''(y)}{\We\, c(y)^2}+2 \beta\,  k^2\,+\mri k_x\, Re\, U(y)+\mri \omega\,  Re,\\ 
a_{1,11} \;=&\; -\frac{4 \mri (1-\beta)\,  k_x\, c''(y) U'(y)}{\We\, c(y)^3}+\frac{8 (1-\beta)\,  k_x^2\, T_{xy}(y) c'(y) U'(y)}{c(y)^3}+\frac{12 (1-\beta)\,  k_x^2\, c'(y) U'(y)^2}{\We\, c(y)^4}\\
&\;-\frac{2 (1-\beta)\,  k_x^2\, c'(y)}{\We\, c(y)^2}+\frac{2 \mri (1-\beta)\,  k_x\, c'(y) T_{xy}'(y)}{c(y)^2}-\frac{8 \mri (1-\beta)\,  k_x\, c'(y) U''(y)}{\We\, c(y)^3}\\
&\;+\frac{12 \mri (1-\beta)\,  k_x\, c'(y)^2 U'(y)}{\We\, c(y)^4}-\frac{2 (1-\beta)\,  k_z^2\, c'(y)}{\We\, c(y)^2}+\frac{2 \mri (1-\beta)\,  k^2\, k_x\, T_{xy}(y)}{c(y)}\\
&\;-\frac{2 \mri (1-\beta)\,  k_x^3\, T_{xx}(y) U'(y)}{c(y)^2}-\frac{4 \mri (1-\beta)\,  k_x^3\, T_{xy}(y) U'(y)^2}{c(y)^3}+\frac{(1-\beta)\,  k_x^2\, T_{xx}'(y)}{c(y)}\\
&\;-\frac{2 (1-\beta)\,  k_x^2\, T_{xy}'(y) U'(y)}{c(y)^2}-\frac{4 (1-\beta)\,  k_x^2\, T_{xy}(y) U''(y)}{c(y)^2}-\frac{8 (1-\beta)\,  k_x^2\, U'(y) U''(y)}{\We\, c(y)^3}, \\ 
a_{0,11} \;=&\; \frac{(1-\beta)\,  k^2\, c''(y)}{\We\, c(y)^2}+\frac{(1-\beta)\,  k_x^2\, T_{xx}(y) c''(y)}{c(y)^2}+\frac{4 (1-\beta)\,  k_x^2\, T_{xy}(y) c''(y) U'(y)}{c(y)^3}+\frac{\mri (1-\beta)\,  k_x\, c''(y) T_{xy}'(y)}{c(y)^2}\\
&\;-\frac{2 \mri (1-\beta)\,  k^2\, k_x\, T_{xy}(y) c'(y)}{c(y)^2}+\frac{4 \mri (1-\beta)\,  k^2\, k_x\, c'(y) U'(y)}{\We\, c(y)^3}-\frac{2 (1-\beta)\,  k^2\, c'(y)^2}{\We\, c(y)^3}\\
&\;+\frac{4 \mri (1-\beta)\,  k_x^3\, T_{xx}(y) c'(y) U'(y)}{c(y)^3}+\frac{12 \mri (1-\beta)\,  k_x^3\, T_{xy}(y) c'(y) U'(y)^2}{c(y)^4}+\frac{(1-\beta)\,  k_x^2\, c'(y) T_{xx}'(y)}{c(y)^2}\\
&-\frac{2 (1-\beta)\,  k_x^2\, T_{xx}(y) c'(y)^2}{c(y)^3}+\frac{4 (1-\beta)\,  k_x^2\, c'(y) T_{xy}'(y) U'(y)}{c(y)^3}+\frac{8 (1-\beta)\,  k_x^2\, T_{xy}(y) c'(y) U''(y)}{c(y)^3}\\
&\;-\frac{12 (1-\beta)\,  k_x^2\, T_{xy}(y) c'(y)^2 U'(y)}{c(y)^4}-\frac{2 \mri (1-\beta)\,  k_x\, c'(y)^2 T_{xy}'(y)}{c(y)^3}-\frac{(1-\beta)\,  k^4\,}{\We\, c(y)}-\frac{(1-\beta)\,  k^2\, k_x^2\, T_{xx}(y)}{c(y)}\\
&\;-\frac{2 (1-\beta)\,  k^2\, k_x^2\, T_{xy}(y) U'(y)}{c(y)^2}+\frac{\mri (1-\beta)\,  k^2\, k_x\, T_{xy}'(y)}{c(y)}-\frac{2 \mri (1-\beta)\,  k^2\, k_x\, U''(y)}{\We\, c(y)^2}-\frac{2 \mri (1-\beta)\,  k_x^3\, T_{xx}'(y) U'(y)}{c(y)^2}\\
&\;-\frac{2 \mri (1-\beta)\,  k_x^3\, T_{xx}(y) U''(y)}{c(y)^2}-\frac{4 \mri (1-\beta)\,  k_x^3\, T_{xy}'(y) U'(y)^2}{c(y)^3}-\frac{8 \mri (1-\beta)\,  k_x^3\, T_{xy}(y) U'(y) U''(y)}{c(y)^3}\\
&\;-\frac{2 (1-\beta)\,  k_x^2\, T_{xy}'(y) U''(y)}{c(y)^2}-\beta  k^4\,-\mri k^2\, k_x\, Re\, U(y)-\mri k^2\, \omega \, Re-\mri k_x\, Re\, U''(y),\\ 
a_{2,21} \;=&\; -\frac{\mri (1-\beta)\,  k_z\, U'(y)}{\We\, c(y)^2},\\ 
a_{1,21} \;=&\; \frac{\mri (1-\beta)\,  k_z\, T_{xy}(y) c'(y)}{c(y)^2}+\frac{4 \mri (1-\beta)\,  k_z\, c'(y) U'(y)}{\We\, c(y)^3}+\frac{3 (1-\beta)\,  k_x\, k_z\, T_{xy}(y) U'(y)}{c(y)^2}\\
&\;+\frac{4 (1-\beta)\,  k_x\, k_z\, U'(y)^2}{\We\, c(y)^3}-\frac{2 \mri (1-\beta)\,  k_z\, U''(y)}{\We\, c(y)^2},
    \end{align*}
    \begin{align*}
      a_{0,21} \;=&\; -\frac{(1-\beta)\,  k_x\, k_z\, T_{xx}(y) c'(y)}{c(y)^2}-\frac{4 (1-\beta)\,  k_x\, k_z\, T_{xy}(y) c'(y) U'(y)}{c(y)^3}-\frac{\mri (1-\beta)\,  k_z\, c'(y) T_{xy}'(y)}{c(y)^2}\\
      &\;+\frac{2 \mri (1-\beta)\,  k_x^2\, k_z\, T_{xx}(y) U'(y)}{c(y)^2}+\frac{4 \mri (1-\beta)\,  k_x^2\, k_z\, T_{xy}(y) U'(y)^2}{c(y)^3}+\frac{\mri (1-\beta)\,  k_x^2\, k_z\, U'(y)}{\We\, c(y)^2}\\
      &\;+\frac{2 (1-\beta)\,  k_x\, k_z\, T_{xy}(y) U''(y)}{c(y)^2}+\frac{\mri (1-\beta)\,  k_z^3\, U'(y)}{\We\, c(y)^2}+\mri k_z\, Re\, U'(y),\\ 
      a_{2,22} \;=&\; -\frac{(1-\beta)\, }{\We\, c(y)}-\beta,\\ 
      a_{1,22 } \;=&\; \frac{(1-\beta)\,  c'(y)}{\We\, c(y)^2}-\frac{2 \mri (1-\beta)\,  k_x\, T_{xy}(y)}{c(y)}-\frac{\mri (1-\beta)\,  k_x\, U'(y)}{\We\, c(y)^2},\\ 
      a_{0,22} \;=&\; \frac{\mri (1-\beta)\,  k_x\, T_{xy}(y) c'(y)}{c(y)^2}+\frac{(1-\beta)\,  k^2\,}{\We\, c(y)}+\frac{(1-\beta)\,  k_x^2\, T_{xx}(y)}{c(y)}+\frac{(1-\beta)\,  k_x^2\, T_{xy}(y) U'(y)}{c(y)^2}\\
      &\;-\frac{\mri (1-\beta)\,  k_x\, T_{xy}'(y)}{c(y)}+\beta  k^2\,+\mri k_x\, Re\, U(y)+\mri \omega\,  Re. 
    \end{align*}
    The operators $\MM C$ (for the velocity output) and $\MM B$ are given by~\cite{jovbamJFM05}
\begin{equation}
  \MM C = \frac{1}{k^2}\left[\begin{array}{cc}
    \mri\,k_x\,\mathrm D & - \mri\,k_z\\
  k^2 & 0 \\
  \mri\,k_z\,\DD & \mri\,k_x
  \end{array}\right],\quad \MM B = \left[\begin{array}{ccc}
    -\mri\,k_x\,\DD & - k^2 & - \mri\,k_z\,\DD\\
    \mri\,k_z & 0 & - \mri\,k_x
  \end{array}\right].
  \end{equation}
  For the stress output $\tau_{xx}$, $\MM C$ is a $1\times 2$ block-matrix operator with
  \begin{align}\label{eq:Ctxx}
    \MM C_{11} \;=\; \sum_{n \, = \, 0}^2 c_{n,11}(y,\omega)\DD^n,
    ~~
    \MM C_{12} \;=\; \sum_{n \, = \, 0}^1 c_{n,12}(y,\omega)\DD^n,
  \end{align}
  where,
    \begin{align*}
      c_{2,11} \;=&\; \frac{2 \mri k_x\, T_{xy}(y)}{k^2\, c(y)}+\frac{2 \mri k_x\, U'(y)}{k^2\, \We\, c(y)^2},\\ 
c_{1,11} \;=&\; -\frac{2 k_x^2\, T_{xx}(y)}{k^2\, c(y)}-\frac{2 k_x^2\,}{k^2\, \We\, c(y)}+\frac{2 k_z^2\, T_{xy}(y) U'(y)}{k^2\, c(y)^2}+\frac{4 U'(y)^2}{\We\, c(y)^3},\\ 
c_{0,11} \;=&\; \frac{2 \mri k_x\, T_{xx}(y) U'(y)}{c(y)^2}+\frac{4 \mri k_x\, T_{xy}(y) U'(y)^2}{c(y)^3}+\frac{2 \mri k_x\, U'(y)}{\We\, c(y)^2}-\frac{T_{xx}'(y)}{c(y)}-\frac{2 T_{xy}'(y) U'(y)}{c(y)^2},\\ 
c_{1,12} \;=&\; -\frac{2 \mri k_z\, T_{xy}(y)}{k^2\, c(y)}-\frac{2 \mri k_z\, U'(y)}{k^2\, \We\, c(y)^2},\\ 
c_{0,12} \;=&\; \frac{2 k_x\, k_z\, T_{xx}(y)}{k^2\, c(y)}+\frac{2 k_x\, k_z\, T_{xy}(y) U'(y)}{k^2\, c(y)^2}+\frac{2 k_x\, k_z\,}{k^2\, \We\, c(y)}.
    \end{align*}
    
    	\vspace*{-4ex}
    \subsection{Descriptor form with the stress eliminated}\label{sec:appCDes}
    
    \vspace*{-2ex}
    The state variables are velocity and pressure, $\bphi = [\,u ~ v ~ w ~ p\,]^T$ in~\eqref{eq:regularOp}, and the boundary conditions are 
    \begin{align*} 
        u(\pm 1) \, = \, v(\pm 1) \, = \, w(\pm 1) \, = \, [\DD v(\cdot)](\pm 1) \, = \, 0.
      \end{align*}
    The relations in \eqref{eq:tauRels} can be used to eliminate stress from \eqref{eq:2.1}. In this representation the operator-valued matrix $\MM A$ is of size $4 \times 4$ and its components are given by
\noindent \begin{align*}
  \MM A_{ij} \; = \; \sum_{n \, = \, 0}^2 a_{n,ij}(y,\omega)\DD^n,
\end{align*}
where the non-zero coefficients $a_{n,ij}$ are given by
\begin{align*}
  a_{2,11} \;=&\; -\frac{(1-\beta)\, }{\We \, c(y)}-\beta,\\ 
 a_{1,11} \;=&\; \frac{(1-\beta)\,  \left(c'(y)-\mri k_x\, \left(3 \We \, c(y) T_{xy}(y)+2 U'(y)\right)\right)}{\We \, c(y)^2},\\ 
 a_{0,11} \;=&\; \frac{(1-\beta)\,  k_x\, T_{xy}(y) \left(2 k_x\, U'(y)+\mri c'(y)\right)}{c(y)^2}+\frac{(1-\beta)\,  \left(2 k_x^2+k_x\, \We \, \left(2 k_x\, T_{xx}(y)-\mri T_{xy}'(y)\right)+k_z^2\right)}{\We \, c(y)}\\
 &\;+\beta  k^2+\mri k_x\, Re\, U(y)+\mri \omega\,  Re,\\
 a_{2,12} \;=&\;  -\frac{(1-\beta)\,  \left(\We \, c(y) T_{xy}(y)+2 U'(y)\right)}{\We \, c(y)^2},\\ 
 a_{1,12} \;=&\; \frac{(1-\beta)\,  T_{xy}(y) c'(y)}{c(y)^2}+\frac{4 (1-\beta)\,  c'(y) U'(y)}{\We \, c(y)^3}-\frac{\mri (1-\beta)\,  k_x\, T_{xx}(y)}{c(y)}-\frac{4 \mri (1-\beta)\,  k_x\, T_{xy}(y) U'(y)}{c(y)^2}\\
 &\;-\frac{4 \mri (1-\beta)\,  k_x\, U'(y)^2}{\We \, c(y)^3}-\frac{\mri (1-\beta)\,  k_x\,}{\We \, c(y)}-\frac{2 (1-\beta)\,  U''(y)}{\We \, c(y)^2},\\ 
 a_{0,12} \;=&\; \frac{\mri (1-\beta)\,  k_x\, T_{xx}(y) c'(y)}{c(y)^2}+\frac{4 \mri (1-\beta)\,  k_x\, T_{xy}(y) c'(y) U'(y)}{c(y)^3}+\frac{\mri (1-\beta)\,  k_x\, c'(y)}{\We \, c(y)^2}-\frac{(1-\beta)\,  c'(y) T_{xy}'(y)}{c(y)^2}\\
 &\;+\frac{2 (1-\beta)\,  k_x^2 T_{xx}(y) U'(y)}{c(y)^2}+\frac{4 (1-\beta)\,  k_x^2 T_{xy}(y) U'(y)^2}{c(y)^3}+\frac{2 (1-\beta)\,  k_x^2 U'(y)}{\We \, c(y)^2}\\
 &\;-\frac{2 \mri (1-\beta)\,  k_x\, T_{xy}(y) U''(y)}{c(y)^2}+\frac{(1-\beta)\,  k_z^2 U'(y)}{\We \, c(y)^2}+Re\, U'(y),\\
 a_{1,13} \;=&\; -\frac{\mri (1-\beta)\,  k_z \left(\We \, c(y) T_{xy}(y)+U'(y)\right)}{\We \, c(y)^2},\\ 
 a_{0,13} \;=&\; \frac{(1-\beta)\,  k_x\, k_z \left(\We \, c(y) T_{xx}(y)+c(y)+\We \, T_{xy}(y) U'(y)\right)}{\We \, c(y)^2},\\
 a_{0,14} \;=&\; \mri k_x\,,\\
 a_{1,21} \;=&\; -\frac{\mri (1-\beta)\,  k_x\,}{\We \, c(y)},\\ 
 a_{0,21} \;=&\; \frac{(1-\beta)\,  k_x^2 T_{xy}(y)}{c(y)},\\ 
a_{2,22} \;=&\; -\frac{2 (1-\beta)\, }{\We \, c(y)}-\beta,\\
a_{1,22} \;=&\; \frac{2 (1-\beta)\,  c'(y)}{\We\, c(y)^2}-\frac{3 \mri (1-\beta)\,  k_x\, T_{xy}(y)}{c(y)}-\frac{2 \mri (1-\beta)\,  k_x\, U'(y)}{\We\, c(y)^2},\\ 
 a_{0,22} \;=&\; \frac{2 \mri (1-\beta)\,  k_x\, T_{xy}(y) c'(y)}{c(y)^2}+\frac{(1-\beta)\,  k_x^2 T_{xx}(y)}{c(y)}+\frac{2 (1-\beta)\,  k_x^2 T_{xy}(y) U'(y)}{c(y)^2}\\
 &\;+\frac{(1-\beta)\,  k_x^2}{\We\, c(y)}-\frac{\mri (1-\beta)\,  k_x\, T_{xy}'(y)}{c(y)}+\frac{(1-\beta)\,  k_z^2}{\We\, c(y)}+\beta  k^2+\mri k_x\, Re\, U(y)+\mri \omega\,  Re,\\
 a_{1,23} \;=&\; -\frac{\mri (1-\beta)\,  k_z}{\We \, c(y)},\\ 
 a_{0,23} \;=&\; \frac{(1-\beta)\,  k_x\, k_z T_{xy}(y)}{c(y)},\\ 
\end{align*}
\begin{align*}
 a_{1,24} \;=&\; 1,\\ 
 a_{0,31} \;=&\; \frac{(1-\beta)\,  k_x\, k_z}{\We \, c(y)},\\ 
 a_{1,32} \;=&\; -\frac{\mri (1-\beta)\,  k_z}{\We \, c(y)},\\ 
 a_{0,32} \;=&\; \frac{(1-\beta)\,  k_z \left(k_x\, U'(y)+\mri c'(y)\right)}{\We \, c(y)^2},\\ 
a_{2,33} \;=&\; -\frac{(1-\beta)\, }{\We \, c(y)}-\beta,\\ 
 a_{1,33} \;=&\; \frac{(1-\beta)\,  c'(y)}{\We \, c(y)^2}-\frac{2 \mri (1-\beta)\,  k_x\, T_{xy}(y)}{c(y)}-\frac{\mri (1-\beta)\,  k_x\, U'(y)}{\We \, c(y)^2},\\ 
 a_{0,33} \;=&\; \frac{\mri (1-\beta)\,  k_x\, T_{xy}(y) c'(y)}{c(y)^2}+\frac{(1-\beta)\,  k_x^2 T_{xx}(y)}{c(y)}+\frac{(1-\beta)\,  k_x^2 T_{xy}(y) U'(y)}{c(y)^2}\\
 &\;+\frac{(1-\beta)\,  k_x^2}{\We \, c(y)}-\frac{\mri (1-\beta)\,  k_x\, T_{xy}'(y)}{c(y)}+\frac{2 (1-\beta)\,  k_z^2}{\We \, c(y)}+\beta  k^2+\mri k_x\, Re\, U(y)+\mri \omega\,  Re,\\
 a_{0,34} \;=&\; \mri k_z.\\
\end{align*}

  \end{subequations}
  The expressions for $\MM B$ and $\MM C$ are given by
  \begin{equation*}
    \MM B 
    \; = \, 
    \left[\begin{array}{ccc}
      1 & 0 & 0\\
      0 & 1 & 0 \\
      0 & 0 & 1\\
      0 & 0 & 0
    \end{array}\right],
    ~~
    \MM C 
    \; = \, 
    \left[\begin{array}{cccc}
      1 & 0 & 0 & 0\\
      0 & 1 & 0 & 0\\
      0 & 0 & 1 & 0
    \end{array}\right],
    \end{equation*}
and for the stress output $\tau_{xx}$, $\MM C$ is a $1\times 3$ block-matrix operator given by
\begin{align}
  \MM C_{ij} \;=&\; \sum_{n \, = \, 0}^1 c_{n,ij}(y,\omega)\DD^n,\label{eq:3DtxxDesH}
\end{align}
where the non-zero coefficients $c_{n,ij}$ are given by
\begin{align*}
  c_{1,11} \;=&~ \frac{2 \left(\We\, c(y) T_{xy}(y)+U'(y)\right)}{\We\, c(y)^2},\\ 
  c_{0,11} \;=&~ \frac{2 \mri k_x\, \left(\We\, c(y) T_{xx}(y)+c(y)+\We\, T_{xy}(y) U'(y)\right)}{\We\, c(y)^2},\\ 
  c_{1,12}  \;=&~ \frac{2 U'(y) \left(\We\, c(y) T_{xy}(y)+2 U'(y)\right)}{\We\, c(y)^3},\\ 
  c_{0,12} \;=&~ \frac{2 \mri c(y) U'(y) \left(k_x\, \We\, T_{xx}(y)+k_x\,+\mri \We\, T_{xy}'(y)\right)-\We\, c(y)^2 T_{xx}'(y)+4 \mri k_x\, \We\, T_{xy}(y) U'(y)^2}{\We\, c(y)^3}.
\end{align*}

\end{document}